\documentclass[preprint]{aastex631}

\usepackage{hyperref}
\usepackage{color}
\usepackage{soul}
\usepackage{natbib}
\usepackage{hyperref}
\usepackage{amsmath,bm}
\usepackage{xspace}
\usepackage{graphicx}
\usepackage{enumitem}
\usepackage{amssymb}
\usepackage{xifthen}
\usepackage{hyperref}
\usepackage[normalem]{ulem}
\usepackage{comment}
\usepackage{amsmath}

\hypersetup{    
  colorlinks      = {true},
  linkcolor       = {blue},
  citecolor       = {blue},
  urlcolor        = {blue},
}

\graphicspath{{./}{figures/}}

\definecolor{orange}{rgb}{1.0,0.5,0.}

\DeclareMathOperator{\sech}{sech}

\def\MDM{\ifmmode{\>M_{\textnormal{\sc dm}}}\else{$$M_{\textnormal{\sc dm}}}\fi}

\def\XH{\ifmmode{\>X_{\textnormal{\sc h}}} \else{$X_{\textnormal{\sc h}}$}\fi}
\def\nH{\ifmmode{\>n_{\textnormal{\sc h}}} \else{$n_{\textnormal{\sc h}}$}\fi}

\def\maspyr{\ifmmode{\>\textnormal{mas~yr}^{-1}}\else{mas~yr$^{-1}$}\fi}

\def\mG{\ifmmode{\>\mu\mathrm{G}}\else{$\mu$G}\fi}
\def\erg{\ifmmode{\> {\rm erg}}\else{erg}\fi}
\def\keV{\ifmmode{\> {\rm keV}}\else{keV}\fi}

\def\deg{\ifmmode{\>^{\circ}}\else{$^{\circ}$}\fi}
\def\onedeg{\ifmmode{\>1^{\circ}}\else{$1^{\circ}$}\fi}

\def\xvir{\ifmmode{\>\!x_{vir}}\else{$x_{vir}$}\fi}
\def\Mvir{\ifmmode{\>\!M_{vir} }\else{$M_{vir} $}\fi}
\def\rvir{\ifmmode{\>\!r_{vir}}\else{$r_{vir}$}\fi}
\def\vvir{\ifmmode{\>\!v_{vir}}\else{$v_{vir}$}\fi}
\def\Vvir{\ifmmode{\>\!V_{vir} }\else{$V_{vir} $}\fi}

\def\tratio{\ifmmode{\>\tau}\else{$\tau$}\fi}

\def\rms{\ifmmode{\>r_{\textnormal{\sc ms}}}\else{$r_{\textnormal{\sc ms}}$}\fi}

\def\Mpc{\ifmmode{\>\!{\rm Mpc}} \else{Mpc}\fi}
\def\kpc{\ifmmode{\>\!{\rm kpc}} \else{kpc}\fi}
\def\pkpc{\ifmmode{\>\!{\rm kpc}^{-1}} \else{kpc$^{-1}$}\fi}
\def\pc{\ifmmode{\>\!{\rm pc}} \else{pc}\fi}

\def\Gyr{\ifmmode{\>\!{\rm Gyr}} \else{Gyr}\fi}
\def\Myr{\ifmmode{\>\!{\rm Myr}} \else{Myr}\fi}
\def\yr{\ifmmode{\>\!{\rm yr}} \else{yr}\fi}
\def\pyr{\ifmmode{\>\!{\rm yr}^{-1}}\else{yr$^{-1}$} \fi}
\def\s{\ifmmode{\>\!{\rm s}}\else{s}\fi}
\def\ps{\ifmmode{\>\!{\rm s}^{-1}}\else{s$^{-1}$}\fi}
\def\Hz{\ifmmode{\>\!{\rm Hz}}\else{Hz}\fi}

\def\kms{\ifmmode{\>\!{\rm km\,s}^{-1}}\else{km~s$^{-1}$}\fi}

\def\K{\ifmmode{\>\!{\rm K}}\else{K}\fi}

\def\sr{\ifmmode{\>\!{\rm sr}}\else{sr}\fi}
\def\psr{\ifmmode{\>\!{\rm sr}^{-1}}\else{sr$^{-1}$}\fi}
\def\arcs{\ifmmode{\>\!{\rm arcsec}}\else{arcsec}\fi}
\def\parcs{\ifmmode{\>\!{\rm arcsec}^{-1}}\else{arcsec${-1}$}\fi}
\def\parcss{\ifmmode{\>\!{\rm arcsec}^{-2}}\else{arcsec${-2}$}\fi}

\def\cm{\ifmmode{\>\!{\rm cm}}\else{cm}\fi}
\def\cc{\ifmmode{\>\!{\rm cm}^{3}}\else{cm$^{3}$}\fi}
\def\sqc{\ifmmode{\>\!{\rm cm}^{2}}\else{cm$^{2}$}\fi}
\def\pcc{\ifmmode{\>\!{\rm cm}^{-3}}\else{cm$^{-3}$}\fi}
\def\psc{\ifmmode{\>\!{\rm cm}^{-2}}\else{cm$^{-2}$}\fi}

\def\g{\ifmmode{\>\!{\rm g}}\else{g}\fi}
\def\Msun{\ifmmode{\>\!{\rm M}_{\odot}}\else{M$_{\odot}$}\fi}
\def\hMsun{\ifmmode{\> h^{-1}{\rm M}_{\odot}}\else{$h^{-1}$M$_{\odot}$}\fi}

\def\Zsun{\ifmmode{\>\!{\rm Z}_{\odot}}\else{Z$_{\odot}$}\fi}

\def\Lsun{\ifmmode{\>\!{\rm L}_{\odot}}\else{L$_{\odot}$}\fi}

\def\rayl{\ifmmode{\>\!{\rm R}}\else{R}\fi}
\def\mR{\ifmmode{\>\!{\rm mR}}\else{mR}\fi}

\def\lya{\ifmmode{\>\!{\rm Ly}\alpha}\else{Ly$\alpha$}\fi}

\def\Ha{\ifmmode{\>\!{\rm H}\alpha}\else{H$\alpha$}\fi}
\def\Hb{\ifmmode{\>\!{\rm H}\beta}\else{H$\beta$}\fi}

\def\HI{\ifmmode{\> \textnormal{\ion{H}{i}}} \else{\ion{H}{i}}\fi}
\def\HII{\ifmmode{\> \textnormal{\ion{H}{ii}}} \else{\ion{H}{ii}}\fi}
\def\CIV{\ifmmode{\> \textnormal{\ion{C}{iv}}} \else{\ion{C}{iv}}\fi}
\def\SiIV{\ifmmode{\> \textnormal{\ion{S}{iv}}} \else{\ion{Si}{iv}}\fi}

\def\NH{\ifmmode{\> {\rm N}_{\rm H}} \else{N$_{\rm H}$}\fi}
\def\Ng{\ifmmode{\> {\rm N}_{\rm gas}} \else{N$_{\rm gas}$}\fi}
\def\NHI{\ifmmode{\> {\rm N}_{\HI}} \else{N$_{\HI}$}\fi}
\def\MHI{\ifmmode{\> {\rm M}_{ \HI}} \else{M$_{\HI}$}\fi}

\def\mua{\ifmmode{\>\mu_{ \textnormal{\Ha}}}\else{$\mu_{ \textnormal{\Ha}}$}\fi}
\def\alphabha{\ifmmode{\>\alpha_{B}^{(\textnormal{\Ha})}}\else{$\alpha_{B}^{(\textnormal{\Ha})}$}\fi}

\newcommand{\ramses}{{\sc Ramses}}
\newcommand\agama{{\sc agama}}

\defcitealias{pla20a}{Planck (2020)}

\shorttitle{Barred galaxies at high redshift}
\shortauthors{Bland-Hawthorn, Tepper-Garc\'ia, Agertz \& Freeman}

\begin{document}

\title{The rapid onset of stellar bars in the baryon-dominated centres of disk galaxies}


\author[0000-0001-7516-4016]{Joss~Bland-Hawthorn}
\affiliation{Sydney Institute for Astronomy, School of Physics, A28, The University of Sydney, NSW 2006, Australia}
\affiliation{Centre of Excellence for All-Sky Astrophysics in Three Dimensions (ASTRO 3D), Australia}

\author[0000-0001-7516-4016]{Thor Tepper-Garcia}
\affiliation{Sydney Institute for Astronomy, School of Physics, A28, The University of Sydney, NSW 2006, Australia}
\affiliation{Centre of Excellence for All-Sky Astrophysics in Three Dimensions (ASTRO 3D), Australia}

\author[0000-0002-4287-1088]{Oscar Agertz}
\affiliation{Lund Observatory, Department of Astronomy and Theoretical Physics, Lund University, Box 43, SE-221 00 Lund, Sweden}

\author[0000-0001-6280-1207]{Ken Freeman}
\affiliation{Research School of Astronomy \& Astrophysics, Mount Stromlo Observatory, Woden, ACT 2611, Australia}
\affiliation{Centre of Excellence for All-Sky Astrophysics in Three Dimensions (ASTRO 3D), Australia}


\correspondingauthor{J. Bland-Hawthorn}
\email{jonathan.bland-hawthorn@sydney.edu.au}
 
\begin{abstract}
Recent observations of high-redshift galactic disks ($z \approx 1-3$) show a strong negative trend in the dark matter fraction $f_{\rm DM}$ with increasing baryon surface density. For this to be true, the inner baryons must dominate over dark matter in early massive galaxies, as observed in the Milky Way today. If disks are dominant at early times, we show that stellar bars form promptly within these disks, leading to a high bar fraction. New James Webb Space Telescope observations provide the best evidence for mature stellar bars in this redshift range. The disk mass fraction $f_{\rm disk}$ within $R_s = 2.2 R_{\rm disk}$ is the dominant factor determining how rapidly a bar forms. Using 3D hydro simulations of halo-disk-bulge galaxies, we confirm the ``Fujii relation" for the exponential dependence of the bar formation time $\tau_{bar}$ as a function of $f_{\rm disk}$. For $f_{\rm disk} > 0.3$, the bar formation time declines exponentially fast with increasing $f_{\rm disk}$. Instead of Fujii’s arbitrary threshold for when a bar appears, for the first time, we exploit the exponential growth timescale associated with the positive feedback cycle as the bar emerges from the underlying disk. A modified, mass-dependent trend is observed for halos relevant to systems at cosmic noon ($10.5 < \log M_{\rm halo} < 12$), where the bar onset is slower for higher mass halos at a fixed $f_{\rm disk}$. If baryons dominate over dark matter within $R \approx R_s$, we predict that a high fraction of bars will be found in high-redshift disks long before $z = 1$.

\end{abstract}

\section{Introduction} \label{s:intro}

There are now independent observations of galaxy disk components in place as early as $z\approx 6$ \citep{riz22w,nee21c} and maybe even earlier
\citep{alg22z,Tokuoka2022}. This is contrary to what is seen in many contemporary cosmological simulations where disks with high rates of rotational velocity appear to form at much later times \citep[e.g.][]{feng15,zav16o,dub16r,pil18l,ferreira22}.
How these disks assemble, form and evolve are pressing issues in astrophysics \citep{fre02d,som15n}.

\vspace{0.1cm}
\noindent {\sl Early disks.}
Mature stellar disks are well established over the redshift range $z\approx 1-3$, both with and without stellar bulges.
Supporting evidence includes Adaptive Optics (AO)-assisted, integral field spectrograph observations of ionized gas \citep{gen06h,for18c}, Hubble Space Telescope (HST) observations of lens-reconstructed disks \citep{Johnson2017,yua17d,dun19j}, Atacama Large Millimeter/sub-millimeter Array (ALMA) observations of [C~II] and molecular disks \citep{tsu21u,lel21s} and, most recently, first light observations from the JWST \citep[][]{guo22}. These objects
appear mostly to be rotation-dominated, massive galaxies, with $V_c/\sigma_{\rm los} \sim 10$, where $V_c$ and $\sigma_{\rm los}$ are the inferred circular velocity and internal velocity dispersion along the line of sight. \cite{rob22} demonstrate that sources from the earlier HST Cosmic Assembly Near-infrared Deep Extragalactic Legacy Survey (CANDELS) labelled as compact are disk-like in JWST Near Infrared Camera (NIRCAM) observations, with over 90\% of sources in their study having a Sersic index $n < 2$ associated with ``disky'' systems. Most recently, \cite{kartaltepe23} has shown that 60\% of galaxies in the redshift range $z=3-6$ are disk systems, with about 30\% being disky in the interval $z=6-9$.

\vspace{0.1cm}
\noindent {\sl Early bars and spiral arms.}
Recently, \cite{guo22} provided spectacular evidence for well-developed stellar bars in the redshift range $z=1.1 - 2.3$ that look much like their mature low-redshift counterparts. These observations are remarkable in several ways. First, the bars are most prominent in the restframe near-infrared band (F444W filter centred at 4.4 microns) and much less so in restframe optical bands. This may explain why earlier papers based on Hubble Space Telescope (HST) observations (mostly limited to the optical restframe) argued for bars being scarce at this epoch relative to low-redshift samples \citep{sim14k}. 

Secondly, the first six galaxies extracted from the Cosmic Evolution Eearly Release Science (CEERS) survey fields \citep{fin22} have baryonic masses ($M_\star \approx 10^{10-11}$ M$_\odot$) and star formation rates ($20-300$ M$_\odot$ yr$^{-1}$) associated with high gas fractions and dispersion velocities ($30-100$ km s$^{-1}$).
These ``hot disks'' are associated with strong gas turbulence that is thought to suppress or delay the formation of bars \citep{ath86,ath03,she12}. Interestingly, spiral arms have also been observed in turbulent (``hot") disk galaxies at $z\approx 2-2.5$ \citep{daw03a,Law12,mar22}. These observations suggest that a common mechanism like swing amplification ($\S4.1$) is able to operate in the presence of high gas fractions and turbulent gas (but see $\S4.5$).

But not all early disks are dominated by star formation and highly turbulent gas. \cite{fud22} identify a significant population of disks in the $z=1-3$ redshift range with red restframe colours \citep[see also][]{nel22n}. These appear to have low or non-existent star formation rates, and may or may not retain gas. Contrary to recent claims \cite[e.g.][]{fud22}, red disks are relatively common in the local Universe and, like the high-redshift sample, may comprise a post-starburst disk population, with or without gas retention \citep{darriush16,kennedy16}.

\vspace{0.1cm}
\noindent {\sl Nuclear regions and central cusps.}
Of paramount importance is how the galaxy's nuclear gravitational potential assembled at early times.  Dark matter simulations with different initial density fluctuation spectra and cosmological parameters, but without baryons, indicate consistently that the dark halos are dominated by central cusps where the dark matter density profile undergoes a rapid upturn towards the nucleus \citep[][NFW]{nav96a}.  The popular NFW model has a density profile $\rho_{\rm DM} \propto (r/r_s)^{-1} (1+r/r_s)^{-2}$ where $r_s$ is the radial scale factor.

But these extraordinary cusps are not observed in disk galaxies today\footnote{How the concentration parameter $c$, halo mass $M_{\rm halo}$ and virial radius $R_{\rm vir}$ ($= c ~r_s$) may have evolved for the Milky Way are shown in Fig. 1 of \cite{bla16a}.}: baryons entirely dominate the inner 20 kpc (diameter) of Milky Way and M31-like galaxies \citep[e.g.][]{bla16a,eil19a}. Intriguingly, these cusps are not even observed in massive disk galaxies at $z\approx 1-2.5$ \citep{wuy16,gen20,Price2021}. The latter papers have established an inverse relation between the dark matter fraction $f_{\rm DM}$ and the baryon surface density $\Sigma_{\rm bary}$ or, conversely, the baryon fraction $f_{\rm bary}$ increases as $\Sigma_{\rm bary}$ increases. If we correct for the maximal contribution from a central bulge, the disk mass fraction $f_{\rm disk}$ tends to increase as the baryon surface density increases, at a fixed halo mass. As we show ($\S$3), these are ideal conditions for triggering the onset of a stellar bar, with the onset being more rapid as $\Sigma_{\rm bary}$ increases.

So what happened to the cusps?
A popular explanation is that baryonic inflows and outflows early in the galaxy's evolution \citep[e.g.][]{pon12n}, or something more exotic like emergent dark-matter annihilation \citep[e.g.][]{cli21y}, have smoothed out the precursor dark matter cusp. While baryonic processes are effective in the low mass limit, these do not operate well in massive galaxies ($V_c \gtrsim 150$ km s$^{-1}$).
Moreover, star formation and AGN activity were at their peak at an epoch ($z\sim 2-3$) where baryon-dominated disk galaxies are observed to exist.
Interestingly, \citet{wei07z} and \citet{sel08a} argue that the early formation of a bar can transform the cusp into a flattened core within $1-2$ Gyr, although this claim has been challenged \citep[][]{dub09b}. But again, there may be no `cusp-core' problem if baryons overwhelmed the nuclear regions at early times. The early formation of bars, once established, can help funnel subsequent baryons to the centre to assist with the baryon domination \citep{ath02}.

The sequence of events that govern the central regions of galaxies bears on, for example, the remarkable `Magorrian relation' establishing a clear relationship between the evolution of the central black hole and the stellar bulge \citep{mag98r,fer00m}. The stellar and gaseous metallicities of massive bulges are solar, or even super-solar, to at least $z\sim 5$ establishing that deep gravitational potentials set up very early on \citep{ham99}. Such metallicities must have been created {\it in situ} and not from the accretion of smaller systems. The nuclear bulge regions are also associated with metal-enriched globular cluster systems, at least in the local Universe \citep{mas19}, indicative of extreme pressures during their formation. 

\vspace{0.1cm}
\noindent{\sl Disk instabilities.} Galactic disk components characterised by large-scale ordered motions are low entropy systems that are acutely sensitive to internal or external perturbations \citep[e.g.][]{ban22}. We can use this fact to explore some of the properties of early disks. For example, we have known since \cite{ost73a} that a bar can be suppressed or delayed by embedding a disk in a dominant DM halo. This is because global disk instabilities are suppressed when the disk contribution to the rotation curve is reduced \citep{too81}. Conversely, this suggests that the disk fraction by mass $f_{\rm disk}$ within some scale radius is a key variable determining the onset of bar formation. More fundamentally, a halo that is massive enough to suppress inner bars is likely to suppress all bisymmetric features like spiral arms or tidally induced perturbations \citep{too81,ath86}. {\it Thus, the detailed structure of early disks is a powerful constraint on the relative contributions of dark matter and baryons during the early stages of galaxy formation.} 

\medskip\noindent
\noindent{\sl Numerical simulations.} Several groups have addressed the formation of disks with bars and spiral arms in cosmological simulations \citep[q.v.][]{rosas22}, which have the advantage of tying disk formation to the assembly history of the system. However, crucially, we point out that the total number of particles per synthetic Milky Way mass galaxy is $N\lesssim 10^{5.5}$ particles, a low-resolution regime that is subject to numerical artefacts \citep{wil22a} and where the vertical structure of the disk is not resolved. These are fundamental concerns if 3D disk perturbations are the focus of the study \citep{ban22}.

With reference to the new JWST observations \citep[e.g.][]{guo22},
we seek to understand when galaxy disks first began to settle as equilibrium configurations in the early Universe. 
Using hydrodynamical, N-body simulations of isolated galaxies with enhanced resolution,
we consider settled disks (gas and stars) that are in equilibrium with the dark matter (DM) halo and a stellar bulge. These halo-bulge-disk (HBD) models are parametrized in terms of total halo mass, disk mass fraction within a fixed diameter, and variable gas fraction with respect to the total disk mass. 

A comprehensive survey of this kind was carried out for the first time by \cite{fuj18a} who explored `dry' (without gas) Milky Way mass simulations with $N \approx 10^7$ disc particles (but with $10^8$ halo particles) to investigate the impact of the changing disk fraction $f_{\rm disk}$ on the bar formation time, $\tau_{\rm bar}$. We confirm but also update some of the key results from this work; we extend to a wider range in $f_{\rm disk}$, and explore the impact of a gaseous disk component with changing $f_{\rm disk}$ and gas fraction $f_{\rm gas}$.
But in our first paper, we consider only low gas fractions ($f_{\rm gas}\leq 0.1$) because a simple equation of state can be used to set up the equilibrium hybrid (gas $+$ stars) system.
In our follow-up paper, we consider the more difficult problem of high
gas fractions ($f_{\rm gas} > 0.5$) where a turbulent medium operates within an HBD set up (Tepper-Garc\'ia et al 2023).

Crucially, we improve on the definition used by \cite{fuj18a} for when a bar has `formed'  by recognizing that the growth of linear instabilities is exponential by nature, particularly when self-excited through a positive feedback loop \citep{bin20}. This is precisely what we see with the initial
growth of the quadrupole moment ($A_2/A_0$) when the bar first appears. Thus we are able to use an exponential growth timescale (e-folding time) rather than the arbitrary use of an $A_2/A_0$ threshold. This step reveals some interesting new behaviour.

The outline of our paper is follows. The new suite of simulations is described in $\S$2 with the results presented in $\S$3. In $\S$4, we discuss these results in light of the high redshift observations and discuss the need for including a turbulent phase and a high gas fraction into the models. We present our conclusions and introduce our future work in $\S$5. Due to its widespread use in simulations, in Appendix A, we investigate the Efstathiou-Lake-Negroponte criterion for monitoring the bar instability. This parameter is found to be inversely related to $f_{\rm disk}$, with a secondary dependence on $M_{\rm halo}$, but is largely unreliable for identifying bars.

\section{Simulations} \label{s:simul}
\subsection{Context} \label{s:cont}

\citet{fuj18a} conducted a series of numerical simulations of a Milky Way-like disk comprising a halo-bulge-disk system. They confirmed two remarkable results that were suggested by
earlier work \citep[e.g.,][]{car85r,val17a}:
\begin{itemize}
    \item The timescale for the onset of the bar instability, $\tau_{\rm bar}$, is mainly governed by $f_{\rm disk}$, the ratio of disk mass to total galaxy mass within the radius at which the rotation curve roughly peaks, where
\begin{equation}
\label{e:fd}
    f_{\rm disk}= \left(\frac{V_{\rm c, disk}(R_s)}{V_{\rm c, tot}(R_s)}\right)_{R_s=2.2 R_{\rm disk}}^2 \, .
\end{equation}
Here
$V_c(R)$ is the circular velocity at a radius $R$, $R_{\rm disk}$ is the exponential disk scalelength, and $R_s=2.2 R_{\rm disk}$ is the traditional scalelength adopted in studies of disks.
    \item {\em All} N-body HBD systems are bar-unstable ultimately. For a sufficiently small value of $f_{\rm disk}$, $\tau_{\rm bar}$ can exceed the age of the Universe.
\end{itemize}

\citet{fuj18a} define $\tau_{bar,0.2}$ as the timespan between the initial state of an unbarred synthetic galaxy, and the epoch at which the maximum value of the normalised quadrupole moment (Fourier $m=2$ mode; see eq.~\ref{e:fourier}) commonly referred to as $A_2/A_0$, crosses the (arbitrary) threshold value of 0.2 (see $\S$3.2). 
In the new work, we adopt Fujii's $A_2/A_0$ criterion, but we also provide an alternative analysis that is less arbitrary and more robust ($\S$3.3).
In addition, they specified an acceptable bar length to be $R_b \gtrsim$ 1 kpc to minimize the impact of resolution on the formation of small inner bars. We have adopted the same threshold in bar length where we use the sudden change in $A_2/A_0$ {\it as a function of radius} to define this quantity. The method used by \citet{fuj18a} was not specified, but we adopt the algorithm described in \cite{dehnen23}. In Fig.~\ref{f:bar}, we show results for model \#7 listed in Tab.~\ref{t:models} that are fairly representative of the models overall. The measured properties from this method are strongly coupled to $A_2/A_0$, as we show.

The upshot is that the higher the value of $f_{\rm disk}$, the shorter the timescale for the onset of the bar instability. For this ratio evaluated at $R=R_s$, \cite{fuj11a} establish that the dividing line occurs at $f_{\rm disk}\approx 0.3$, in the sense that smaller values lead to bar formation timescales that exceed a Hubble time \citep[see also][]{val17a}. They also find that there is an asymptotic limit to the bar formation timescale such that, in the limit of high $f_{\rm disk}$, there is a finite {\it minimum} timescale for bar formation of order 1 Gyr. We investigate both of these claims here.

At cosmic noon, the mass of the host dark matter (DM) halo of Milky Way-type galaxies, $M_{\rm halo}$, was an order of magnitude lower than it is today \citep[][their Fig. 1]{bla16a}. How does $M_{\rm halo}$ affect the dependence of $\tau_{\rm bar}$ on $f_{\rm disk}$? Furthermore, in some disks, the gas fraction was much higher than today, typically gas/disk ratios by mass ($f_{\rm gas}$) of up to 50\%, rather than $\lesssim$10\% associated with nearby disk galaxies today \citep{nar12}.
If $f_{\rm disk}$ includes the contribution of all baryons (stars + gas), and we increase the gas contribution relative to other baryons, does $f_{\rm disk}$ yield the same $\tau_{\rm bar}$ compared to the case without gas? Earlier work suggests that a higher gas fraction reduces the bar's lifetime \citep{vil10k}, but as to formation times, this appears to be unknown.

While \citet{fuj18a} did consider a range of halo masses $M_{\rm halo}$ in their seminal work, they did not explicitly address the dependence of $\tau_{\rm bar}$ on $f_{\rm disk}$ for different $M_{\rm halo}$. A further caveat of their study is that it is restricted to synthetic galaxies without any gas component. Here we explore the impact of different $M_{\rm halo}$ on the bar formation time scale (at fixed $f_{\rm disk}$), as well as the impact of adding a gaseous disk with a low gas fraction $f_{\rm gas}$, relative to the total disk mass, i.e. 
\begin{equation}
\label{e:fdd}
    f_{\rm gas}=
    \left(\frac{M_{\rm disk, gas}}{M_{\rm disk}}\right) ,
\end{equation}
where $M_{\rm disk}$ is the total disk mass and $M_{d, \rm gas}$ is the gas contribution.
We leave further investigation of bar lifetimes in turbulent gas-rich disks to a later paper.

\subsection{Model grid} \label{s:imp}

\begin{table*}
\begin{center}
\begin{tabular}{cccccccccllc}
\hline\\
$M_{\rm halo}$ &  $z$  &  $c$  &  $R_{vir}$ &  $r_{\rm s}$ &  $M_b$ &  $r_b$ &  $M_{\rm disk}$ &  $R_{\rm disk}$ &  $f_{\rm disk}$  &  Label (identifier) & $f_{\rm gas}$ \\
($10^{10}$~\Msun) &  &  &  (kpc) &  (kpc) &  ($10^{4}$~\Msun) &  (kpc) & ($10^{8}$~\Msun) &  (kpc) &  &  & \\
\hline\\
5  &  5  &  3.36  &  19.64  &  5.84  &  3.0  &  0.20  &  10.4  &  0.7  &  0.3  &  Low mass (1)  & -- \\
5  &  5  &  3.36  &  19.64  &  5.84  &  3.0  &  0.20  &  25.0  &  1.0  &  0.35  &  Low mass (19)  & -- \\
5  &  5  &  3.36  &  19.64  &  5.84  &  3.0  &  0.20  &  29.4  &  1.0  &  0.4  &  Low mass (2)  & -- \\
5  &  5  &  3.36  &  19.64  &  5.84  &  3.0  &  0.20  &  116  &  1.8  &  0.5  &  Low mass (3) & -- \\
5  &  5  &  3.36  &  19.64  &  5.84  &  3.0  &  0.20  &  274  &  2.5  &  0.6  &  Low mass (4)  & -- \\
5  &  5  &  3.36  &  19.64  &  5.84  &  3.0  &  0.20  &  501  &  3.0  &  0.7  &  Low mass (13)  & -- \\
5  &  5  &  3.36  &  19.64  &  5.84  &  3.0  &  0.20  &  921  &  3.0  &  0.8  &  Low mass (14)  & -- \\
10  &  3  &  4.05  &  36.96  &  9.24  &  6.0  &  0.28  &  9.79  &  0.7  &  0.3  &  Medium mass (5)  & -- \\
10  &  3  &  4.05  &  36.96  &  9.24  &  6.0  &  0.28  &  22.5  &  1.0  &  0.35  &  Medium mass (20)  & -- \\
10  &  3  &  4.05  &  36.96  &  9.24  &  6.0  &  0.28  &  29.4  &  1.0  &  0.4  &  Medium mass (6)  & 0.1 \\
10  &  3  &  4.05  &  36.96  &  9.24  &  6.0  &  0.28  &  112  &  1.8  &  0.5  &  Medium mass (7)  & 0.1 \\
10  &  3  &  4.05  &  36.96  &  9.24  &  6.0  &  0.28  &  274  &  2.5  &  0.6  &  Medium mass (8)  & 0.1 \\
10  &  3  &  4.05  &  36.96  &  9.24  &  6.0  &  0.28  &  555  &  3.0  &  0.7  &  Medium mass (15)  & 0.1 \\
10  &  3  &  4.05  &  36.96  &  9.24  &  6.0  &  0.28  &  880  &  3.0  &  0.8  &  Medium mass (16)  & -- \\
50  &  1  &  5.85  &  121.70  &  20.80  &  30  &  0.62  &  7.57  &  0.7  &  0.3  &  High mass (9)  & -- \\
50  &  1  &  5.85  &  121.70  &  20.80  &  30  &  0.62  &  18.7  &  1.0  &  0.35  &  High mass (21)  & -- \\
50  &  1  &  5.85  &  121.70  &  20.80  &  30  &  0.62  &  23.7  &  1.0  &  0.4  &  High mass (10)  & -- \\
50  &  1  &  5.85  &  121.70  &  20.80  &  30  &  0.62  &  103  &  1.8  &  0.5  &  High mass (11)  & -- \\
50  &  1  &  5.85  &  121.70  &  20.80  &  30  &  0.62  &  273  &  2.5  &  0.6  &  High mass (12) & -- \\
50  &  1  &  5.85  &  121.70  &  20.80  &  30  &  0.62  &  499  &  3.0  &  0.7  &  High mass (17) & -- \\
50  &  1  &  5.85  &  121.70  &  20.80  &  30  &  0.62  &  884  &  3.0  &  0.8  &  High mass (18) & --
\end{tabular}
\caption{Overview of galaxy models. Columns are as follows: (1) Host halo virial mass; (2) Galaxy redshift; (3) Host halo concentration; (4) Host halo virial radius; (5) Host halo scale radius; (6) Bulge mass; (7) Bulge scale radius; (8) Stellar disk mass; (9) Stellar disk scalelength; (10) disk-to-total mass ratio (Eq.~\ref{e:fd}); (11) Model designation; (12) Gas to stellar disk mass fraction. Note that for each gas-bearing model there exist a dry counterpart.}
\label{t:models}
\end{center}
\end{table*}

In this study, a synthetic galaxy is modelled as a three-component system consisting of a host DM halo, a bulge, and a stellar disk, sampled with 2, 0.5, and 1 million particles, respectively. In addition to this {\em dry} model, we consider or four-component extended model that includes a gaseous disk, i.e. a {\em gas-bearing} galaxy. The gas is modelled with an adaptive mesh that tracks of order $10^7$ independent grid points (see $\S 2.3$).

In order to simulate redshift evolution, we focus on the halo/stellar mass in our isolated simulations because many studies have found that mass and environment are the dominant factors in galaxy evolution \citep[e.g.][]{peng10}. These factors determine the accretion rate over the lifetime of the galaxy. The next most important factors are the disc mass fraction $f_{\rm disk}$ and the gas mass fraction $f_{\rm gas}$ explored in our study.

With this set up, our goal is to calculate the evolution of both dry and gas-bearing galaxies spanning a wide range in $M_{\rm halo}$, $f_{\rm disk}$ and $f_{\rm gas}$, and to measure $\tau_{\rm bar}$ for each. Thus we construct galaxies with $M_{\rm halo} \in \{ 5\times10^{10}, ~10^{11}, ~5\times10^{11} \}$ \Msun, and $f_{\rm disk} \in \{0.3, ~0.35, ~0.4, ~0.5, ~0.6, ~0.7, ~0.8\}$, yielding 21 collisionless models. In addition, we consider variations of the medium-mass model ($M_{\rm halo} = 10^{11} \Msun$), by adding a gaseous disk with a gas-to-stellar disk fraction (by mass) $f_{\rm gas}  = 0.1$ for the models with $f_{\rm disk} \in \{0.4, ~0.5, ~0.6, ~0.7\}$. Thus, our grid consists of 21 dry models and 4 gas-bearing models.

Our choice of halo mass is motivated by the cosmological growth of a MW-like galaxy, over a broad range in mass, from redshift $z \approx 5$ to $z \approx 1$ \citep[cf.][their fig. 1]{bla16a}.
For self-consistency, the concentration for each halo given its mass is calculated for the corresponding redshift with help of the {\sc commah} package \citep{cor15h}. Thus our galaxy models probe the evolution of a MW-like galaxy over the past 12 to 6 billion years, albeit in a simplistic way.
Our approach complements the different approach of cosmological simulations: these treat the build-up of mass, but have insufficient resolution to study bar formation reliably. As for cosmological simulations, we stress that our dark halos are ``live'' rather than rigid or static; this is a crucial distinction in light of \cite{sel16} who showed that unrealistic static halos slow down the growth rate of the bar.

The properties of each model are summarised in Tab.~\ref{t:models}. The choice of halo mass together with the target $f_{\rm disk}$ value fully dictate the mass of the disk. For a given mass, the corresponding scalelength is estimated with the relation presented by \citet{fat10n}. In practice, we fit a line to their data (see their fig. 15). Given our use of the \agama\ package ($\S 2.3$) to set up dynamically self-consistent HBD models, it requires some trial and error\footnote{The composite model is set up via iteration: one can specify the desired target density profile for each component but the actual profiles will differ slightly
from their target because all components affect each other. The summed potential-density pairs must settle
to a self-consistent configuration.} to arrive at a reasonably self-consistent set of values [$f_{\rm disk}, M_{\rm disk}, R_{\rm disk}$] for a given halo mass, $M_{\rm halo}$.

\begin{figure*}[!htb]
    \centering
    \includegraphics[width=0.9\textwidth]{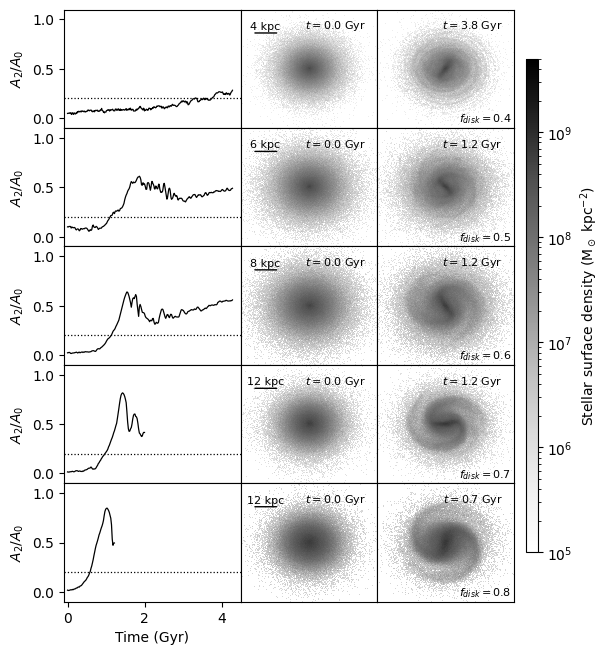}
    \caption{Overview of the mid-mass halo models' evolution for increasing values of the disk-to-total mass ratio $f_{\rm disk}$ (top to bottom). The behaviour of other halo mass models is fairly similar and they are omitted here. Left: Evolution of the bar strength ($A_2/A_0$). The horizontal dotted line indicates the {\em conventional} bar formation threshold ($A_2/A_0 \equiv 0.2$). Note how the timespan for the rapid increase in $A_2 / A_0$ up to its first peak decreases with $f_{\rm disk}$.
    Centre / right: Surface density maps of the stellar disk on a face-on projection at the beginning of the simulation (centre) and roughly 100 Myr after the bar formation threshold has been crossed (right). Note the difference in spatial scale across rows, indicated on the top-left corner of each central panel. The corresponding value of $f_{\rm disk}$ is indicated on the bottom-right corner of each panel on the right column. }
    \label{f:maps}
\end{figure*}

Finally, we add to each of our models a stellar bulge component with a negligible mass to ensure stability in the nuclear disk region (Table 1). This choice is imposed upon us by our present lack of knowledge about the existence and properties of high-redshift bulges. 
It is relatively easy to add significant bulges and these are considered by \cite{fuj18a}. The bulge mass is another free parameter in the simulations but it does not deflect from the importance of $f_{\rm disk}$: the
bar tends to be suppressed where the bulge dominates the potential, but not exclusively.
The issue of small vs. large bulges (e.g. Milky Way vs. M31) may be associated with the specific accretion history a galaxy experiences. This topic has been explored in many recent simulations that form discs with bulges and bars in a cosmological setting \citep{grand17,blazquez20,bi22,rosas22,cavanagh22,reddish22}.

\begin{figure}[!htb]
    \centering
    \includegraphics[width=0.49\textwidth]{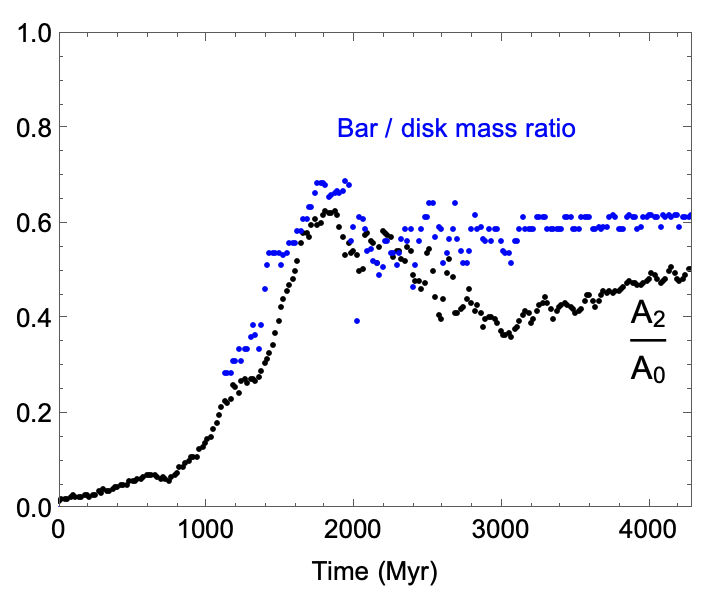}
     \includegraphics[width=0.49\textwidth]{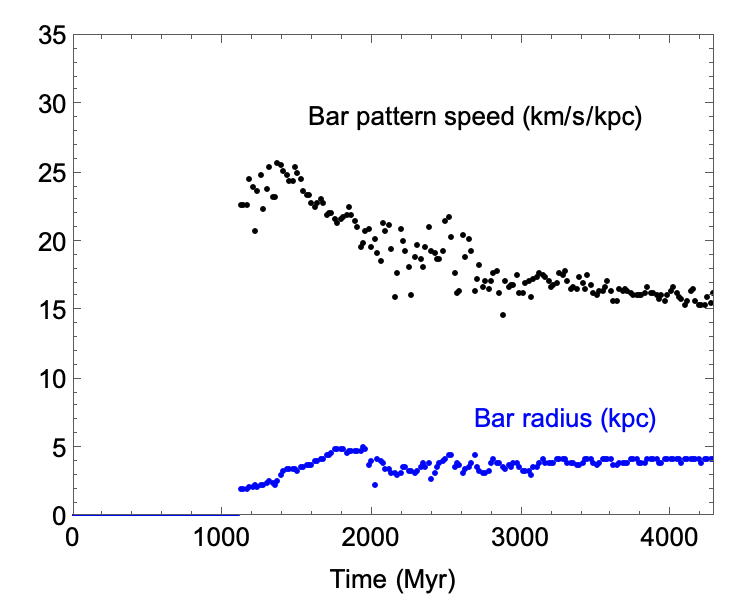}
    \caption{Measured bar parameters as a function of time for model \#7 in Tab.~\ref{t:models}. (Left) In the Dehnen method, the bar's properties are strongly coupled to $A_2/A_0$, as we see here from its comparison to the growth of bar mass to disk mass ratio with time, settling to about 60\% eventually. (Right) The bar pattern speed slows down from 25 to 15 km s$^{-1}$ kpc$^{-1}$ due to the bar's coupling with the live halo; the bar outer radius grows slowly and settles to about 4 kpc.
    }
    \label{f:bar}
\end{figure}

\subsection{Initial conditions and evolution}

The initial conditions for each simulation are generated with the \agama\ package that has been demonstrated to provide reliable initial conditions for long-lived equilibrium disks \citep{vas19a}, with or without gas \citep[cf.][]{bla21e,tep21v,tep22x}.

We have discussed our typical set-up at length in our earlier papers. Here we provide only a brief outline. Pure N-body models are setup exploiting the self-consistent-modelling (SCM) module of \agama. For models with a gas disk, we follow the prescription of \citet{wan10a}, which we have embedded into the SCM module. In brief, the gas disk is strictly isothermal and initially axisymmetric, with a temperature fixed at $T = 10^4$ K to ensure $Q_{\rm gas} > 1$ everywhere. (The Toomre $Q$ stability parameter can be defined for stars or gas using the formula in eq.~\ref{e:Q}.)
It features a surface density profile described by a radially declining exponential, $\Sigma = \Sigma_0 \exp [ -R / R_g]$, with a scale length equal to that of the stellar disk. Its vertical structure follows a $\sech^2 [z/z_0]$ profile, appropriate for a gas distribution in vertical hydrostatic equilibrium, with a scale height $z_0$ that increases with cylindrical radius (a ‘flaring’ disk). The azimuthal velocity profile of the gas disk ensures rotational support against radial instabilities \citep[see][their equation 13]{wan10a}. This prescription is valid only for strictly isothermal gas configurations.

The compound HBD system is evolved with the the N-body/hydrodynamical, Adaptive Mesh Refinement (AMR) \ramses\ code \citep{tey02a}. The system is placed into a cubic box spanning 100 kpc or 200 kpc on a side for the low- and mid-mass halo models and for the high-mass halo models, respectively. The AMR grid is maximally refined up to level 14, implying a limiting spatial resolution of \mbox{$\delta x$ = 100 kpc / $2^{14} \approx 6$ pc} and \mbox{$\delta x$ = 200 kpc / $2^{14} \approx 12$ pc}. The maximum resolution is achieved everywhere within a few disk scalelengths, the most relevant spatial domain of our study.
The mesh is refined if either of these conditions is met: (i) the particle number in any cell exceeds 25; (ii) the gas mass in a cell exceeds $\sim 6 \times 10^4$ \Msun; or (iii) the cell size exceeds the local Jeans length divided by 4 \citep{tru97a}.

Due to the presence of the simulation mesh, the force exerted onto any particle is necessarily softened. In \ramses, the force softening is on the order of a length equivalent to the side of cubic volume element (cell) $\delta x$. In our simulation, this scale is up to an order of magnitude smaller than the typical disk scaleheight, and far below the typical size of the emergent bar.  Hence, in all of our models we resolve well the vertical disk structure, which is essential if we are to follow the evolution of 3D perturbations.
In the models featuring a gas disk, the fluid is evolved subject to a strict isothermal equation of state, whereby the gas is kept at its initial temperature $T = 10^4$ K throughout. Note that with this approach, the gas metallicity becomes irrelevant.

The total simulation time depends on $f_{\rm disk}$, and it ranges from roughly 2 Gyr up to roughly 8 Gyr, for the highest value and lowest $f_{\rm disk}$ values, respectively.

\subsection{Resolution}

We first explore the low $f_{\rm disk}$ limit to ensure that we are not subject to numerical artefacts due to our choice of resolution.
This is an important test because \citet{fuj18a} demonstrate that simulation size (i.e. particle number) and the bar formation timescale are coupled in the `low $N$' limit, in agreement with earlier work \citep[][]{dub09b}.

Moreover, many authors have found that bars can be triggered artificially if the intrinsic resolution is too low, typically $\log_{10} N\lesssim 5.5$ particles \citep{fuj11a,wil22a} in the sense that the emergent bar properties depend on the intrinsic resolution or smoothing length. \cite{gab06} performed tests of bar properties and found general convergence for $\log_{10} N \gtrsim 5.6$ particles.

\cite{fuj11a} specify that $\log_{10} N\gtrsim 6$ particles in the disk is necessary to avoid artificial heating through close encounters of the massive particles, in agreement with others \citep[][]{dub09b,wil22a}.
In our earlier work above, we verified that the
set up for disk simulations does not give rise to spurious disk heating. The vertical dispersion of the particles $\sigma_z(R)$ does not change appreciably at any radius over the 4.3 Gyr timeframe of the simulations. Occasionally, we do find a small amount of heating in the radial dispersion $\sigma_R(R)$ if weak spiral perturbations are created during the disk's evolution. This is not usually addressed by simulators, but this has negligible impact in any event on the rise of the $A_2/A_0$ component.

\subsection{Factors that influence bar formation}

\cite{sel13} has provided a detailed account of the physical and numerical
factors that influence bar formation and the extent of our current understanding. An important non-linear process is `swing amplification' where small perturbations are amplified in a differentially rotating disk \citep{too81}. These generate leading waves and trailing waves that reflect at the disk centre and at corotation. 
A resonant cavity is formed such that the reflecting waves work together to amplify the epicyclic motions of disk particles \citep[for an accessible discussion, see][]{bin08}. 

This mechanism is widely discussed in the context of generating transient spiral arms, but has been implicated in bar formation as well \citep{jul66,too81}.
The amplification factor is given by 
\begin{equation}
    X = \frac{\kappa^2(R) R}{2\pi m\:  G\: \Sigma_{\rm disk}(R)} .
     \label{e:X}
\end{equation}
$\kappa$ is the epicyclic (radial) frequency in the plane of the disk with surface density $\Sigma_{\rm disk}$ at a given radius $R$. $G$ is the gravitational constant and the integer $m$ indicates the specific azimuthal mode that is excited, for which we adopt $m=2$ for bar and bisymmetric spiral modes. 
Note the inverse dependence of the modal amplification $X$ on $\Sigma_{\rm disk}$, which seems counter to the strong dependence of the bar instability on $f_{\rm disk}$.  However, there are other factors that come into play. 

\cite{ath84} showed that the swing amplification factor also depends on the shear rate where
\begin{equation}
    \Gamma = -\frac{d\:\rm ln\: \Omega(R)}{d\:{\rm ln} R}
     \label{e:G}
\end{equation}
for which $\Omega(R)$ is the angular frequency at a radius $R$. $\Gamma = 1$ corresponds to a flat rotation curve with smaller or larger values indicating a rising or falling rotation curve, respectively.
A lower value of $m$ is excited by a higher shear rate \citep{fuj18a}.

Furthermore, the Toomre $Q$ criterion measures stability against {\it local azimuthal (axisymmetric) perturbations} in the presence of shear from differential rotation and internal radial dispersion $\sigma_R$, such that 
\begin{equation}
    Q = \frac{\kappa(R)\:\sigma_R(R)}{a\: G\: \Sigma_{\rm disk}(R)} \, ,
    \label{e:Q}
\end{equation}
where $\kappa$ is the in-plane epicyclic frequency. Strictly speaking, this relation is only valid (i) for a Gaussian velocity distribution, a condition that is not always satisfied in disk galaxies, and (ii) for a flattened distribution, whereas our models include stars, gas and dark matter defined by a wide range of vertical scaleheights.
The constant factor $a=\pi$ for a fluid analysis, and $a=3.36$ for our distribution function approach \citep{too64a}. $Q > 1$ implies local stability of the disk, $Q \sim 1$ indicates marginal or neutral stability, and $Q<1$ is the condition for local instability \citep{saf60,too64a}.

Both parameters $Q$ and $\Gamma$ influence the timescale of the onset of the bar \citep{ath84}, although \cite{fuj18a} find the effects are of secondary importance compared to the primary role of $f_{\rm disk}$. For our models, the shear rate $\Gamma$ is not a free parameter in that the properties of the HBD models are set by a distribution function approach as implemented in \agama, and the reasonable scaling relations we have assumed based on observations ($\S$2.2), although there is room for manoeuvre given the observational uncertainties.

In high-redshift galaxies of particular interest here, the normal assumption is that $Q\sim 1$ to explain the high levels of star formation observed to date \citep{ino16}. This is an obvious selection effect in that these galaxies are most easily studied kinematically. We fix $Q$ across all of our simulations to remove this uncertainty. In light of \cite{ost73a}, \cite{jog14} has extended Eq.~\ref{e:Q} to include the dark halo contribution. 
Thus the stellar disk in our model is characterised by $Q \gtrsim 1.2$ within $2.5 \lesssim R / \kpc \lesssim 10$ to allow instabilities to take hold.

A consequence of fixing $Q$ is that $-$ {\it at fixed halo mass} $-$ as we increase $f_{\rm disk}$, then $\sigma_R$ must increase to compensate for the higher relative surface density. In principle, a higher $\sigma_R$ would make the disk {\it more} stable against bar instabilities \citep{ath86}, which is in the opposite sense of the trend that we see - a higher $f_{\rm disk}$ (producing a higher $\sigma_R$) leads to faster bar formation. This is a consequence of the compensating higher surface density $\Sigma_{\rm disk}$ in Eq.~\ref{e:Q}.

Another factor that influences bar formation is the presence of central bulges. In a photometric study of close to 4000 galaxies, \cite{bar08} finds that bars occur in 60-70\% of late-type bulgeless galaxies, and 40-50\% of early-type disk galaxies. For the latter, the bars are often weak but they are clearly detected in the local universe. This is a likely explanation for why bars are undercounted at higher redshift. A possible mechanism for suppressing bars with dominant bulges is the presence of an inner Lindblad resonance imposed by the rapidly rising central density. These resonances tend to absorb rather than reflect waves excited by swing amplification such that bar formation is suppressed \citep{sel01}. The presence of strong bulges in high-redshift disks is a complicating factor in that they can suppress bars or, when they do exist, render the bars harder to detect.

We do not consider the bulge mass fraction in our study because \cite{fuj18a} has already demonstrated that the disk to {\it total} mass fraction $f_{\rm disk}$
dominates over any dependence on the bulge (mass or scalelength) for situations where the bar is not suppressed by the bulge. An interesting corollary is that if some of the {\it total} mass budget is used to increase the bulge mass at fixed $f_{\rm disk}$, the onset of bar formation is delayed \citep[e.g.][]{sah13}. This delay is quantified in the \cite{fuj18a} study.

\section{Results} \label{s:res}

\subsection{Bar formation and the role of disk mass fraction}

\begin{figure*}[!htb]
\centering
\includegraphics[width=0.32\textwidth]{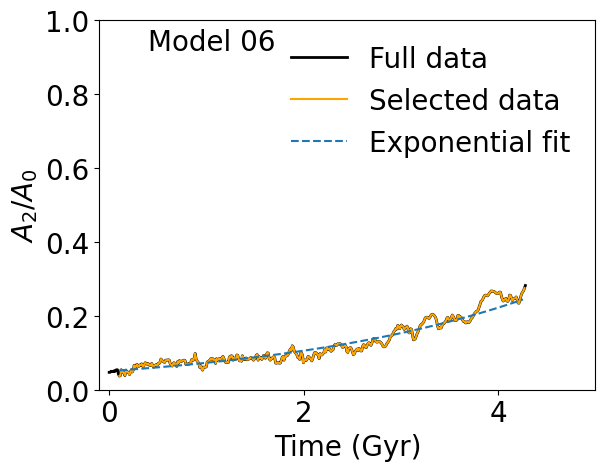}
\includegraphics[width=0.32\textwidth]{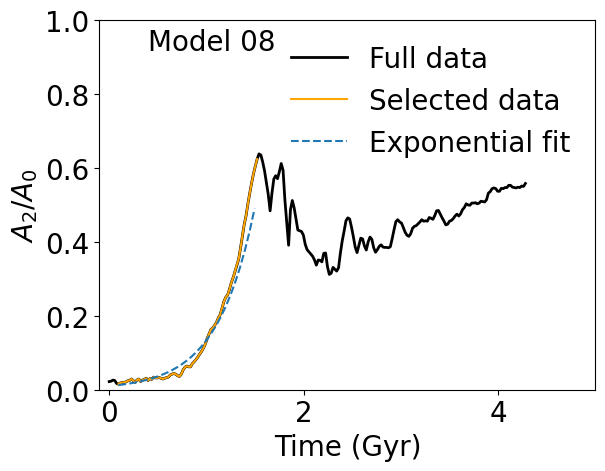}
\includegraphics[width=0.32\textwidth]{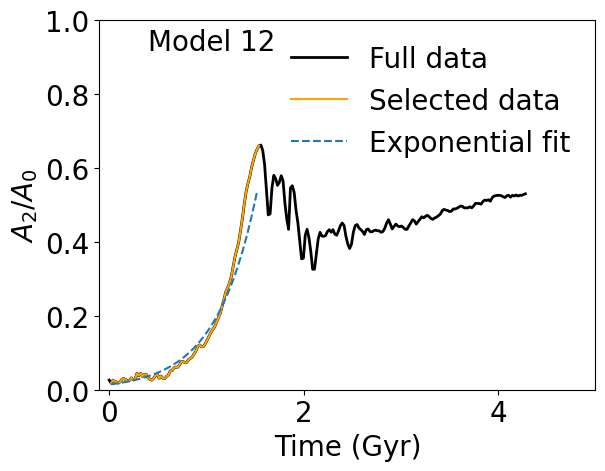}\\
\includegraphics[width=0.32\textwidth]{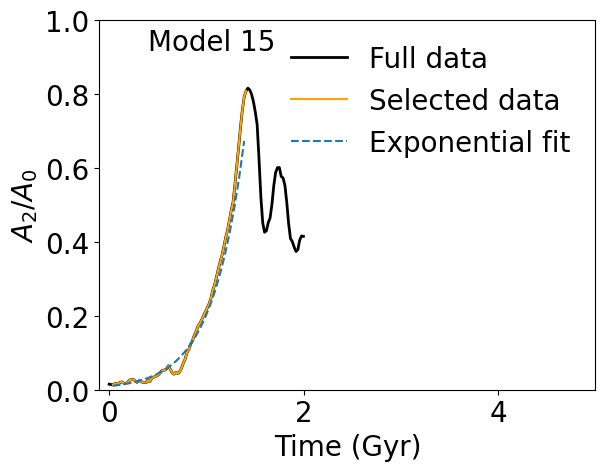}
\includegraphics[width=0.32\textwidth]{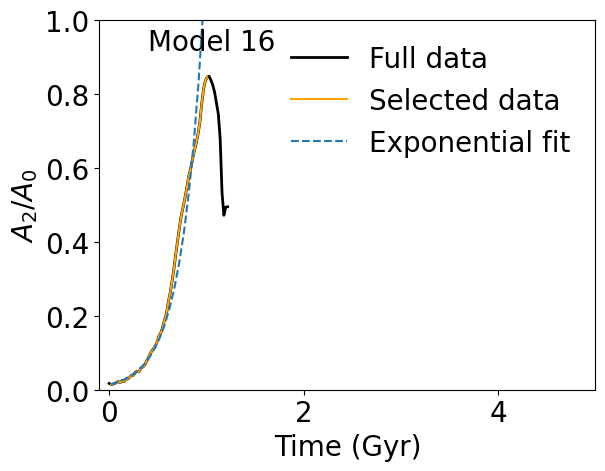}
\includegraphics[width=0.32\textwidth]{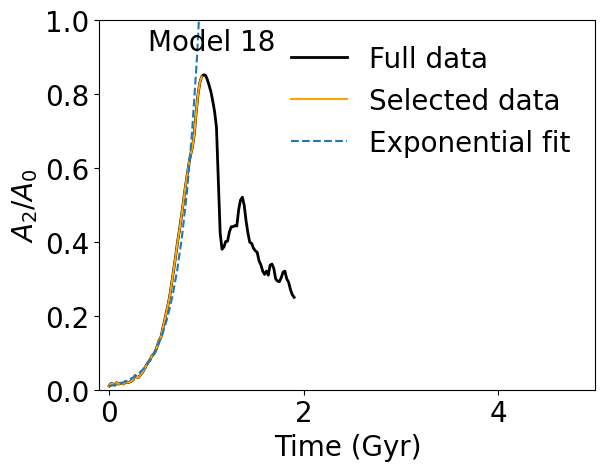}
\caption{Temporal evolution of the quadrupole moment $A_2/A_0$ for a selection of models with different $f_{\rm disk}$. The model is indicated on the top left corner of each panel (Tab.~\ref{t:models}). In each case, the orange overlay indicates the data used to estimate $\tau_{\rm bar}$ using an exponential fit proportional to $\exp[ t / \tau_{\rm bar,exp} ]$, and the blue dashed line indicates the resulting fit. For some models, the runs are terminated shortly after the peak of $A_2/A_0$ to save computer time. }
\label{f:A2}
\end{figure*}

The results of our extensive suite of controlled  three-dimensional bar simulations are shown in Figs.~\ref{f:maps}$-$\ref{f:taubarexp}.
The simulations are available at the following site: 
\url{http://physics.usyd.edu.au/\~tepper/proj\_hizbars.html}.

At each simulated timestep, we carry out a Fourier decomposition of the normalised surface density of the simulated disk, such that
\begin{equation}
\frac{\Sigma(R,\phi)}{\Sigma_0} = \frac{1}{A_0} \sum^{\infty}_{m=0} A_m(R)e^{im[\phi_o-\phi_m(R)]}
    \label{e:fourier}
\end{equation}
for which $A_m(R)$ and $\phi_m(R)$ are the Fourier amplitude and phase angle for the $m$th mode at a radius $R$, and $\Sigma_0$ is the central surface density. In our simulations, $m=2$ is the dominant mode when the bar emerges.

In Fig.~\ref{f:maps} (left panel), we show how the $A_2/A_0$ component of the disk develops with time for five values of $f_{\rm disk}$ using dry, intermediate mass simulations only. In the remaining columns, the equilibrium disk is shown at the start of the simulation (middle) and immediately after the bar has formed (right); note the changing physical scale with halo (disk) mass. The definition of `formed' is the $A_2/A_0$ ($=0.2$) threshold adopted by \cite{fuj18a} as indicated by the dashed line in the first panel. We refer to $\tau_{\rm bar,0.2}$ as the formation time of the bar defined in this way.
The characteristic rise, maximum and settling phase is well established in earlier numerical studies, as already mentioned.

Note that $A_2/A_0$ is initially monotonically increasing for all axisymmetric disk simulations \citep[e.g.][see their Appendix]{tep21v}. 
Thus, regardless of $f_{\rm disk}$, {\it all} disks develop streaming motions in the plane such that they are inherently unstable. It is just that the low $f_{\rm disk}$ cases do not produce a prominent bar in a Hubble time.
In Fig.~\ref{f:A2}, we illustrate how the rise in $A_2/A_0$ to the first peak is well-fitted by an exponential function with a characteristic timescale $\tau_{\rm bar,exp}$.  As our results show below, this provides us with a more robust and less arbitrary way of specifying the timescale for bar formation than adopting a fixed value of $A_2/A_0$. It also means that we do not need to go far beyond the $A_2$ peak, thus saving on expensive computational time, if the goal is to measure this timescale. The exponential rise can be understood in terms of the growth of linear instabilities in a positive feedback cycle; these occur during the `swing amplification' process, which in turn feeds the linear bar modes \citep{too81,sel13,bin20}. At some point, the instabilities become non-linear and the process saturates; some of the power appears in higher order modes.

The movie sequences, in particular, are instructive to view as a function of $f_{\rm disk}$. For example, the 
vertical X-structure long known to occur at the centres of bars \citep[e.g.][]{com90} is particularly prominent in side elevation for the low $f_{\rm disk}$ barred cases, becoming more squashed in high $f_{\rm disk}$ cases, presumably because of the higher disk gravity. 
To our knowledge, this has not been observed before, although \cite{vic20} anticipated such an effect. This X-structure (or peanut-shaped) bulge is commonly observed in edge-on bars with weak bulges, and most notably in the Milky Way \citep[e.g.][]{nes16,por15}.

\subsection{Comparison with earlier work}

\begin{figure*}[!htb]
    \centering
    \includegraphics[width=0.7\textwidth]{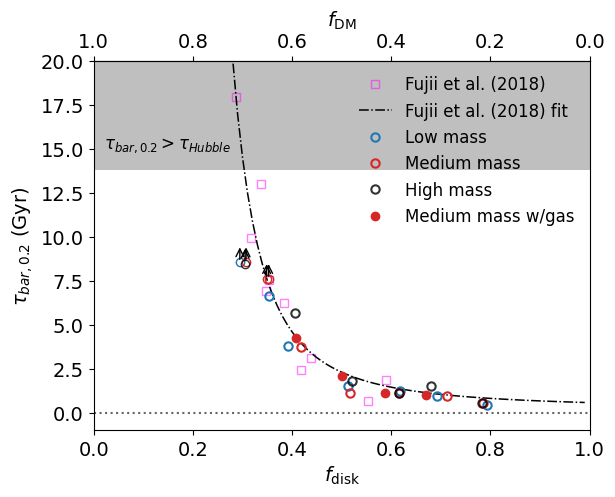}
    \caption{Dependence of the bar formation time $\tau_{bar,0.2}$ on the disk-to-total mass ratio $f_{\rm disk}$ $-$ we refer to this as the `Fujii diagram'. The key indicates the halo mass used, and whether the simulation includes gas or not. The upwards pointing arrows indicate lower limits since the respective simulations were ceased at this time; no bar had formed until that point. The top axis in both panels indicates the dark matter fraction within 2.2 $R_{\rm disk}$ in the absence of a bulge.
    The shaded region indicates where the timescales are longer than a Hubble time, which
    corresponds to $f_{\rm disk} \lesssim 0.3$.
    The data points from the earlier study \citet{fuj18a} are indicated; the dot-dashed curve corresponds to their fit (Eq. \ref{e:fit0}).
    }
    \label{f:taubar}
\end{figure*}

\begin{figure*}[!htb]
    \centering
    \includegraphics[width=0.49\textwidth]{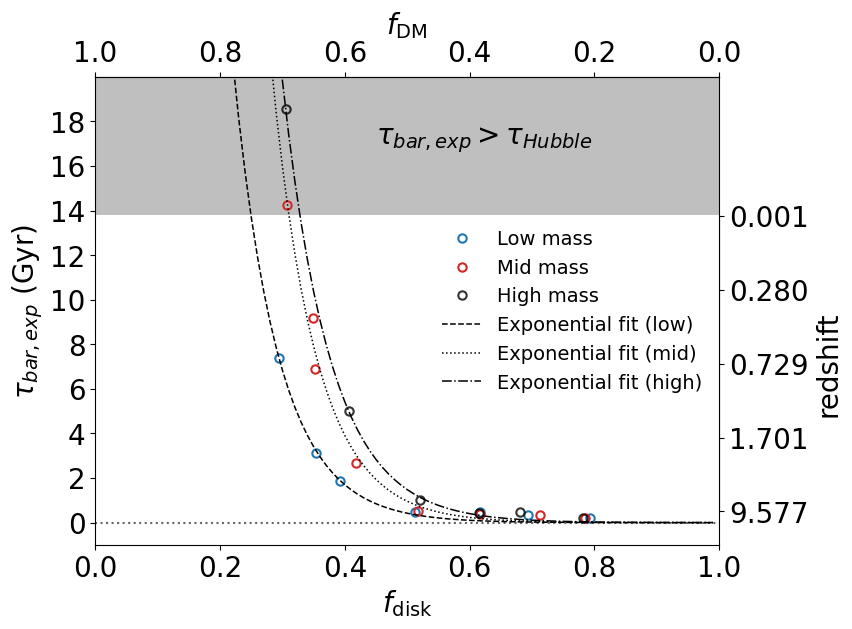}
    \includegraphics[width=0.49\textwidth]{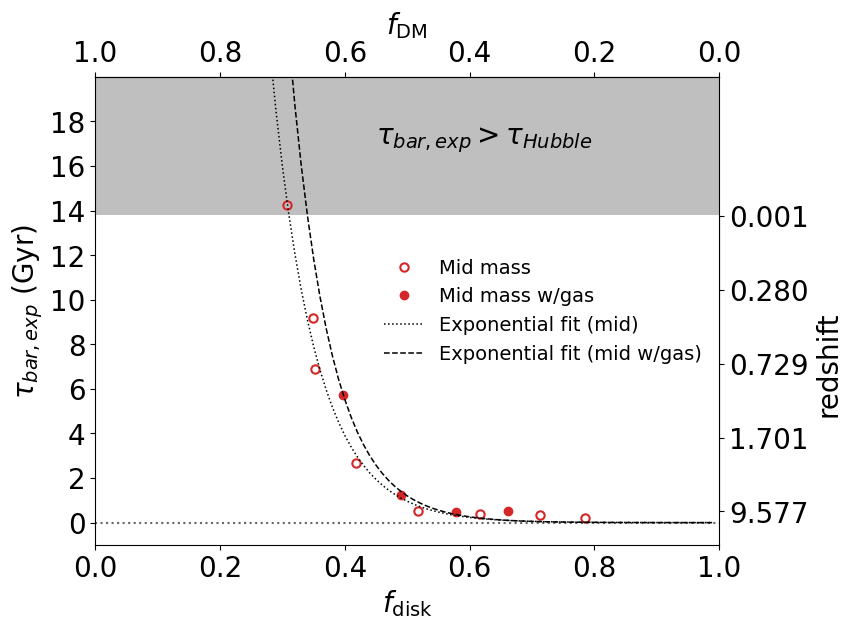}
    \caption{Dependence of the bar formation times $\tau_{bar,exp}$ on the disk-to-total mass ratio $f_{\rm disk}$. For the first time, the timescale is estimated by fitting an exponential function to the rising left flank of the $A_2/A_0$ profile. Left: N-body models for all three halo masses. Right: Mid halo-mass models without gas and with gas ($f_{\rm gas} = 0.1$). Note that the dotted curve with open red circles is repeated in both panels. The top axis in both panels indicates the dark matter fraction within 2.2 $R_{\rm disk}$ in the absence of a bulge. The right axis indicates in each case the bar formation redshift assuming a \citetalias{pla20a} cosmology.  
    }
    \label{f:taubarexp}
\end{figure*}

In Fig.~\ref{f:taubar}, we compare our main findings with \cite{fuj18a}, specifically how the bar formation time $\tau_{\rm bar,0.2}$ depends on $f_{\rm disk}$. The pink boxes are taken directly from the earlier paper. The `Fujii diagram' reveals that the bar formation time drops precipitously for disks with $f_{\rm disk} > 0.3$. This cut-off was first established by \cite{wid08}
but also observed in later work \citet[][see also \citealt{sel14a}]{val17a}.
Fujii et al. only ran their simulations up to $f_{\rm disk}=0.6$ and thus conclude that there is a non-zero asymptotic limit in the bar formation time, i.e a minimum bar formation time of order 1 Gyr. 

In Fig.~\ref{f:taubar}, all other data points refer to our work. Open circles indicate the results for pure N-body models; solid circles show the results for the N-body $+$ gas models with $f_{\rm gas} = 0.1$. The gray shaded area highlights the regime where the bar formation time $\tau_{\rm bar,0.2}$ generally exceeds the age of the Universe.  The dot-dashed curve corresponds to a fit of a declining exponential function and an additive constant established by \cite{fuj18a}, such that
\begin{equation}
    \tau_{\rm bar}/\tau_{\rm H} = 0.0106~\exp[1.38 / f_{\rm disk} ] \, .
    \label{e:fit0}
\end{equation}
where $\tau_{\rm H} \approx 13.78$ Gyr is the Hubble time \citep[]{pla20a}.
This curve agrees well with our data points, which is reassuring: the codes and computer architectures adopted in both studies are very different.

With relevance to our earlier investigation of issues relating to numerical resolution ($\S 2.4$), \cite{fuj18a} demonstrate the impact of inadequate resolution ($N < 10^{5.5}$ particles) on their estimates of $\tau_{\rm bar,0.2}$. The data points (not shown in Fig.~\ref{f:taubar}) fall to the left of the exponential decline in the ``forbidden zone'' shaded in grey. 

With sufficient resolution, the points line up on the curve as shown regardless of resolution scale. We confirm this resolution-dependent behaviour in the new work although find a different trend at high $f_{\rm disk}$, as we discuss below.
The upwards pointing arrows indicate the corresponding values are lower limits. Note that the $f_{\rm disk}$ values for each model have been slightly shifted to avoid symbols overlapping with one another. 
The top horizontal axis indicates the corresponding DM fraction, assuming a negligible bulge contribution within $2.2 R_{\rm disk}$. This aids comparison with the data presented in \cite{Price2021}.

\subsection{New definition for the bar formation timescale}

In Fig.~\ref{f:taubarexp} (left), the same parameter space is shown where $\tau_{\rm bar,0.2}$ is replaced with the exponentially fitted $\tau_{\rm bar,exp}$. This figure only includes our data points because an exponential timescale was not measured by \citeauthor{fuj18a} for their simulations.
We immediately note a relative displacement between the three halo models, such that the trend diverges at low $f_{\rm disk}$. The separate fitted relations are:
\begin{eqnarray}
\tau_{\rm bar}/\tau_{\rm H} &=& \exp[-13.9~(f_{\rm disk} - 0.250)]\ ;\ {\rm low\; mass}
\label{e:fit2}\\
\tau_{\rm bar}/\tau_{\rm H} &=& \exp[-14.1~(f_{\rm disk} - 0.310)]\ ;\ {\rm mid\; mass}
\label{e:fit3}\\
\tau_{\rm bar}/\tau_{\rm H} &=& \exp[-13.0 ~(f_{\rm disk} - 0.328)]\ ;\ {\rm high\; mass}
 \label{e:fit4}
\end{eqnarray}
Note that Fujii's asymptote (minimum bar time $\sim$ 1 Gyr) is no longer seen;
all three relations predict an almost instantaneous bar in the high $f_{\rm disk}$ limit.
Moreover, there is a dependence on halo mass in the sense that higher halo masses have longer bar formation times. This appears to scale with the moderately higher dark matter concentration in the higher mass models. Fujii considered a range of halo masses but does not comment on any specific halo/stellar mass dependence.

In Fig.~\ref{f:taubarexp} (right), for the mid-mass models with gas, the fitted relation is
\begin{equation}
    \tau_{\rm bar}/\tau_{\rm H} = \exp[-15.3 ~(f_{\rm disk} - 0.340)]\ ;\ {\rm mid\,mass/gas}
 \label{e:fit5}
\end{equation}
After dividing Eq.~\ref{e:fit3} by  Eq.~\ref{e:fit5}, it is apparent that, in the low $f_{\rm disk}$ limit ($0.3 < f_{\rm disk} < 0.6$), at a fixed value of $f_{\rm disk}$, the bar formation times are longer by up to 50\%. The presence of gas serves to slow down the onset of the bar. Above that range, within the fit errors, the bar formation times are not significantly altered.
Thus, in the presence of gas, the effect of slower bar formation times is that the models lie above (or equivalently, to the right of) the dashed dividing line in the Fujii diagram (Fig.~\ref{f:taubar}).

\begin{figure}[!htb]
    \centering
    \includegraphics[width=0.49\textwidth]{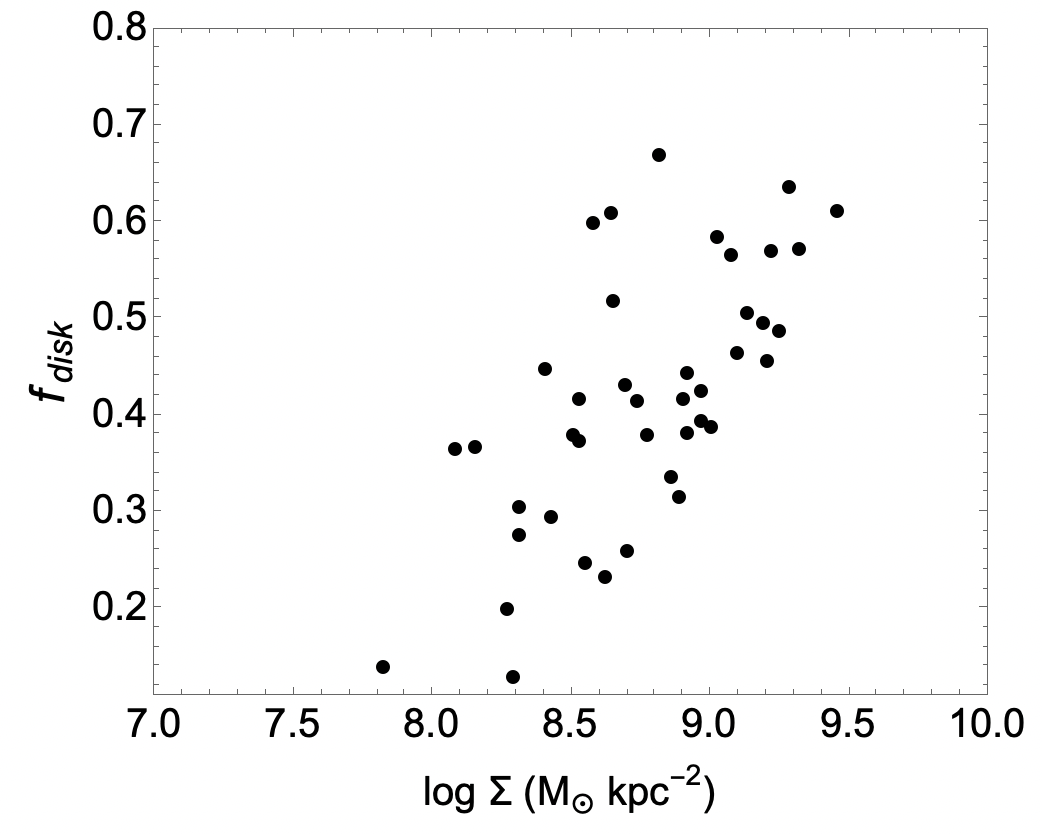}
    \caption{{The inferred values of disk mass fraction $f_{\rm disk}$ taken from \cite{Price2021} after correcting for the quoted bulge fraction. These data points are used in Fig.~\ref{f:pricez}.
    }}
    \label{f:price}
\end{figure}

\section{discussion} \label{s:disk}

\subsection{Implications for high-redshift bars}

Galaxy disks are low-entropy systems with highly ordered rotation, and are highly sensitive to perturbations.
If baryons dominate the inner parts of galaxies for all time, such that dark matter cusps were never able to take hold, we would expect essentially all early disks to exhibit $m=2$ structures, either bars, spiral arms or tidal features. All of these are likely to be enhanced in disks with higher mass fractions.

In Fig.~\ref{f:taubarexp}, in both left and right panels, we relate the bar formation time to cosmic time or redshift. Here we notice something interesting for all of the models. In the high $f_{\rm disk}$ limit ($f_{\rm disk} \gtrsim 0.5$), bars form on timescales of about $100-150$ Myr. Assuming that disks are already forming as early as $z\sim 9$ (see \S 1), bars are likely to be observed to earlier times than the highest-redshift barred system to date \citep[$z\approx 2.3$;][]{guo22}. 
For this to be true, baryons must dominate the centres of galaxies at these earlier epochs \citep{wuy16,gen20,Price2021}. This seems increasingly plausible in light of the discovery of a disk galaxy at $z\approx 4.4$ with well-developed spiral arms \citep{tsu21u}. 

In Fig.~\ref{f:price}, we return to the $z\approx 1-3$ galaxy disk sample presented in \cite{Price2021}. 
The inferred values of the disk mass fraction $f_{\rm disk}$ are plotted against the baryon surface density $\Sigma_{\rm bary}$ after correcting for the quoted bulge fractions. The observed trend simply reflects the fact that a higher surface density in the stars is correlated, on average, with a higher disk mass fraction.

\begin{figure}[!htb]
    \centering
    \includegraphics[width=0.47\textwidth]{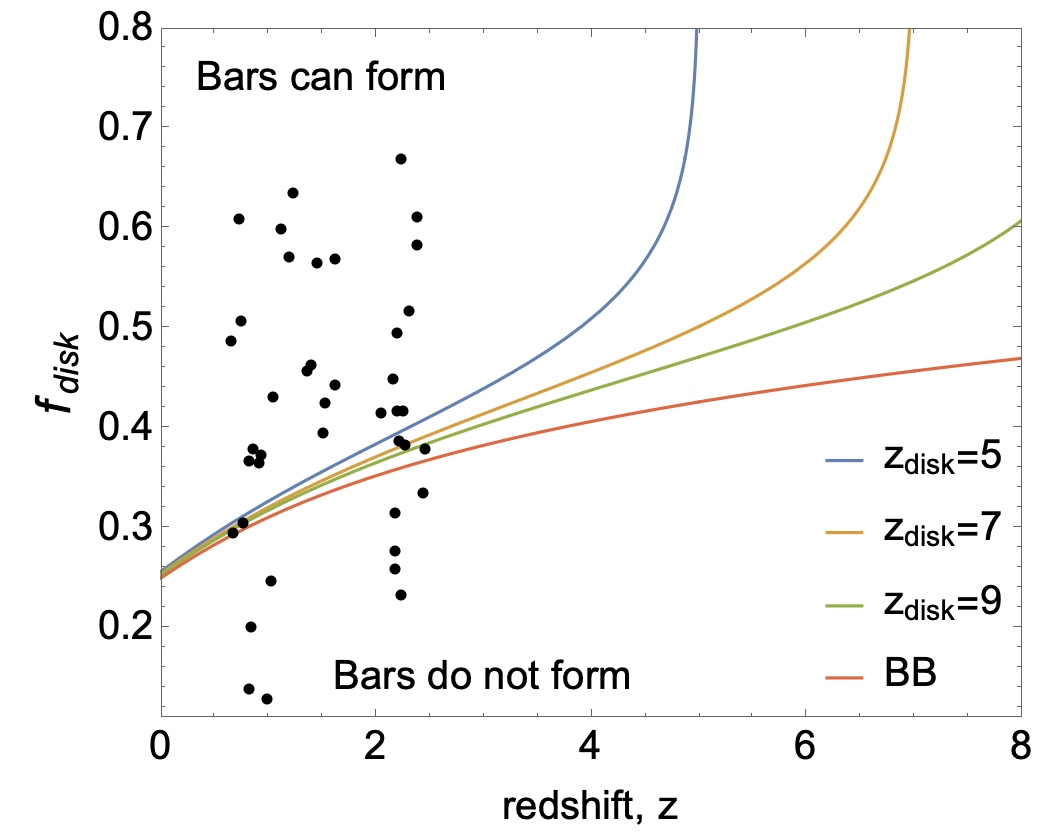}
    \hspace{0.5cm}
     \includegraphics[width=0.47\textwidth]{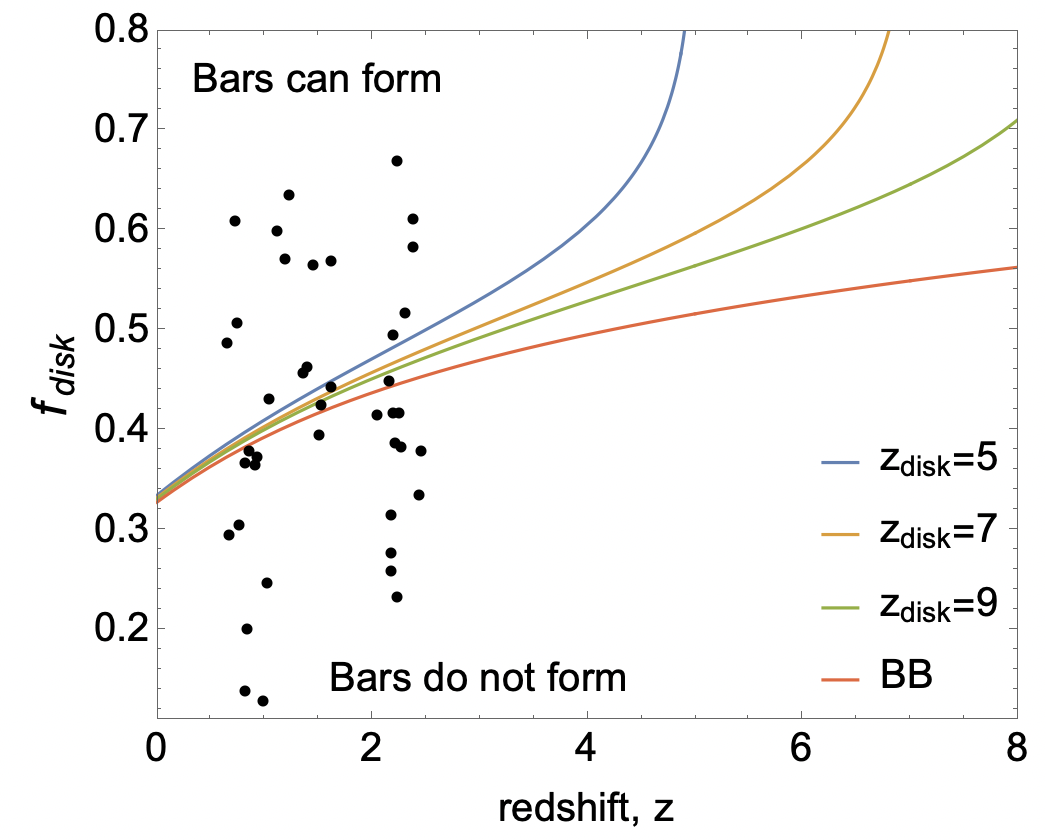}
    \caption{{The inferred values of disk mass fraction $f_{\rm disk}$ shown in Fig.~\ref{f:price} plotted against the measured redshifts for each galaxy. In both figures, the four overlaid curves are the dividing lines between bar formation and insufficient time for bars to form since the epoch of disk formation $z_{\rm disk}$ for four different disk onset times: (blue) $z_{\rm disk}=5$, (yellow) $z_{\rm disk}=7$, (green) $z_{\rm disk}=9$, (red) Big Bang. Disk galaxies that fall above the curves are able to form a bar in the time span available. The left figure, which uses the $f_{\rm disk}(z)$ relation for low mass halos, has 65-70\% of galaxies above the curves. The right figure uses the relation for high mass halos, and has 40-45\% of galaxies above the curves.
    }
    }
    \label{f:pricez}
\end{figure}

The Fujii relation relates the bar onset time to the disk mass fraction. It therefore follows that we can relate the disk mass fraction with cosmic time if we make an assumption about when disks settle for the first time. This is what is illustrated in Fig.~\ref{f:pricez} for four different disk onset times ($z_{\rm disk}=5,7,9,\infty$). In the left and right figures respectively, we equate the timescale of the Fujii relation in the low (eq.~\ref{e:fit2}) and high halo mass limit (eq.~\ref{e:fit4}) offset by the disk formation epoch to the redshift/cosmic time relation in \citet[][eq. 15]{eisenstein97}.

For example, for a redshifted sample at $z=3$, say, the disk mass fraction cut-off to form a bar is higher because there is insufficient time to form bars with disk fractions below $f_{\rm disk}=0.5$. A redshift $z\approx 3$ corresponds to a time 2.2 Gyr after the Big Bang. If disks first settled at $z\approx 9$ consistent with the ALMA data ($\S$1), naively, this gives $\lesssim$2 Gyr for the bar to develop, which requires $f_{\rm disk} \gtrsim 0.5$.

\smallskip
In Fig.~\ref{f:pricez} (left),
the four curves for the low mass limit are as follows:
\begin{eqnarray}
f_{\rm disk}(z) &=& 0.250\, -0.0719424 \ln (-0.0860623 + g(z))\;\;\;\;\;\;\;\;\;\;\;\;\;\;\; z_{\rm disk} = 5 \\
f_{\rm disk}(z) &=& 0.250\, -0.0719424 \ln  (-0.0559584 + g(z))\;\;\;\;\;\;\;\;\;\;\;\;\;\;\; z_{\rm disk} = 7 \\
f_{\rm disk}(z) &=& 0.250\, -0.0719424 \ln (-0.0400557 + g(z))\;\;\;\;\;\;\;\;\;\;\;\;\;\;\; z_{\rm disk} = 9 \\
f_{\rm disk}(z) &=& 0.264\, -0.0719424 \ln (1.220\; g(z))\;\;\;\;\;\;\;\;\;\;\;\;\;\;\;\;\;\;\;\;\;\;\;\;\;\;\;\; z_{\rm disk} = \infty
\end{eqnarray}
where
\begin{equation}
g(z) = 0.819698 \sinh^{-1}\frac{1.54591} {(z+1)^{3/2}}.
\end{equation}
The four curves for the high mass limit in Fig.~\ref{f:pricez} (right) are as follows:
\begin{eqnarray}
f_{\rm disk}(z) &=& 0.328\, -0.0769231 \ln (-0.0860623 + g(z))\;\;\;\;\;\;\;\;\;\;\;\;\;\;\; z_{\rm disk} = 5 \\
f_{\rm disk}(z) &=& 0.328\, -0.0769231 \ln  (-0.0559584 + g(z))\;\;\;\;\;\;\;\;\;\;\;\;\;\;\; z_{\rm disk} = 7 \\
f_{\rm disk}(z) &=& 0.328\, -0.0769231 \ln (-0.0400557 + g(z))\;\;\;\;\;\;\;\;\;\;\;\;\;\;\; z_{\rm disk} = 9 \\
f_{\rm disk}(z) &=& 0.343\, -0.0769231 \ln (1.220\; g(z))\;\;\;\;\;\;\;\;\;\;\;\;\;\;\;\;\;\;\;\;\;\;\;\;\;\;\;\; z_{\rm disk} = \infty
\end{eqnarray}

In Fig.~\ref{f:pricez}, the galaxies observed by the SINFONI integral field spectrograph are shown as points overlaid on the four curves. The low-mass halo models (left) predict that more than two thirds of the sample lie above the curves and therefore may have formed bars. The high-mass halo models (right) predict that less than half will have formed bars.

On the basis of Fig.~\ref{f:pricez}, 
we anticipate future JWST imaging will establish that about half of the Price sample will show evidence of stellar bars. If this statement is found to be true, it will provide independent evidence of (i) established (mature) stellar disks at early times, and (ii) baryon domination over dark matter at the centres of these systems.

\subsection{Future N-body simulations}

A new generation of cosmological simulations of disk formation at higher resolution is called for. \cite{reddish22} show in the {\tt NEWHORIZON} simulations that bar fractions are low (7\%) at $z\sim 1$ and decline to lower redshift. Further investigations reveal that the synthetic galaxies either retain dark matter cusps that dominate over the gas disks, or form dominant stellar bulges that suppress the bars. They outline reasons based on intrinsic sampling and resolution ($\gtrsim 100$ pc) as to why so many related simulations appear to undercount barred galaxies.

In the first cosmological simulations to incorporate baryons, ``overcooling'' was a problem such that baryons easily overwhelmed the dark matter at the centres of galaxies \citep[e.g.][]{ben10a}. Then an armada of feedback prescriptions were introduced to push baryons further out, in order to make disks \citep[e.g.][]{sca08a,age13,grand17,hop20a}. It can be argued that these corrective measures went too far and may account for the suppression of high-redshift disks, particularly disks that dominate the inner potential.

Putting aside the simulations, even in the presence of small to moderate bulges, bars form easily at high disk mass fraction \citep{fuj18a}. 
The disk does not have to be fully settled for this to be true; even thick stellar disks develop bars \citep{gho22}.
Thus confirming that exponential disks exist at early times, and that these disks exhibited strong ordered departures from axisymmetry, is an important clue as to how the inner parts of galaxies formed so quickly, even while the halo was still assembling.

Finally, in light of the problems of numerical resolution \citep[e.g.][]{wil22a}, 
{\it the Fujii diagram (Fig.~\ref{f:taubar}) is a serious challenge to N-body simulators attempting to produce realistic barred galaxies in cosmologically-motivated simulations.}
For the bar formation times not to be compromised by resolution, simulated bars would need to lie on or to the right of the exponential cut-off line in Fig.~\ref{f:taubar}, depending on the specific definition used for the bar timescale.

\subsection{The frequency of barred galaxies}

Numerical studies of the stellar bar phenomenon in galactic dynamics have a long history \citep{com81}, particularly with regard to the bar's role in galaxy evolution \citep{hoh71}. Bars accumulate gas at the galactic centre and fuel nuclear starbursts and encourage the growth of supermassive black holes; star formation is rarely seen along the length of galactic bars \citep{sai22}. Bars transport angular momentum outwards, generate transient spiral arms, and can drive stellar migration. For bars that end at co-rotation, the spiral arms are expected to evolve independently and become detached \citep{sel88}.  Bars appear to trigger small central bulges \citep{com90} as observed in the Milky Way \citep{por15}. Large classical bulges typically suppress bars within the bulge's domain although, in the local universe, bars are observed even in early-type disks \citep{bar08}.

Bars occur in half of all disk galaxies today \citep{bar08,lee19} and so are clearly an important phenomenon to understand. 
In a comprehensive study of 1700 galaxies in the local universe, Lee et al (2019) use three different methods to establish that bars exist in about half of all disks, consistent with many earlier studies (e.g. Eskridge et al 2000), with or without bulges. These numbers may be slightly inflated by their criterion to accept weak as well as strong bar distortions (see also Vera et al 2016). Our simulated bars are uniformly strong in the sense that they account for about 60\% of the inner disk mass within the bar's outer radius (see Fig.~\ref{f:bar}(left)).

But here we are using the presence of a bar as a probe of the disk mass fraction compared to the dark halo. If our central argument is true such that bars only arise where disks (i.e. baryons) dominate over dark matter, the presence of an inner bar indicates that the stellar disk must dominate the local gravity field on the scale of the bar. This argument requires no knowledge of a galaxy's rotation curve. Dark matter halo contributions to galaxy rotation curves are notoriously difficult to get right, even for nearby galaxies, with different authors claiming minimal or maximal disk contributions \citep{kru11}.

In our new simulations, we find that any disk for which $f_{\rm disk} \gtrsim 0.3$ rapidly forms a bar that survives for billions of years, even for a Hubble time in the longest simulations \citep{fuj18a}. 
Based on recent observations of high disk mass fractions at $z=1-3$ \citep{Price2021}, our expectation is that barred disks should be common at this epoch, even within this narrow time window. This assumes the disks are not dominated by massive central bulges. It is unlikely that we currently know how many early disks are barred, with or without bulges, particularly when the observed fraction increases in rest-frame infrared observations compared to optical wavelengths. Early claims for bar fractions declining with increasing redshift were subsequently shown to be incorrect \citep[for a review, see][]{bar08}. 

\cite{bar08} shows that some two thirds of massive late-type disks appear to have bars today, with bars in less than half of early type disks.
If all massive disks dominate their dark matter haloes out to $2.2 R_e$ for all time, then an interesting question is why bars do not exist in all (late-type) disk galaxies today. 
It is unclear whether the problem lies in the formation or dissolution (destruction) of central bars. \cite{ath05} argue that a central mass concentration can destroy a bar under special circumstances. It may be that non-barred disks today had bars in the past \citep{kor04} but that raises the issue of what causes bar transience. \cite{bou02} argue for episodic gas accretion as a natural mechanism for destroying and restoring the bar, primarily through the action of the stellar bar being overloaded with gas, or through the rapid development of a central mass concentration.

A complicating factor is the presence of dominant stellar bulges that would tend to suppress axisymmetric instabilities.
If baryons do take hold in the nuclei of early disk galaxies, 
then it is not certain these are ruled by ordered rotational motions \citep{kre21,gur22}. Large bulges can form through the accretion of large clumps via dynamical friction or dissipative interaction \citep{nog99,zol15}. But the first JWST NIRCAM observations show initial evidence for disks at cosmic noon ($z=1-3$) with a hint of spiral structure and small to moderate stellar bulges \citep{fud22}. Clearly more and deeper imaging is called for, particularly in restframe infrared bands where bars are more evident in the local universe.

\subsection{Bar and spiral structure development in gas rich, high redshift galaxies}

The strong evidence for well-developed stellar bars in massive disk galaxies in the redshift range $z = 1.1-2.3$ \citep{guo22} is a challenge for current theoretical models. High redshift galaxies are observed to be more gas-rich compared to local  galaxies \citep[e.g.][]{gen15,sai16,tac18}, as well as featuring higher levels of gas turbulence \citep[e.g.][]{wis15x}. Using the near-infrared integral field spectrograph SINFONI, rapidly star-forming high-redshift galaxies were found to feature gas velocity dispersions as high as $100$ km s$^{-1}$ \citep{epi09,gen11}. At face value, such conditions are thought to suppress the formation of bars \citep{ath86,she12}, in conflict with the recent JWST observations.

Our understanding of the ISM conditions in high redshift galaxies has improved greatly in the past decade. Star forming disks in the KMOS$^{\rm 3D}$ integral field spectroscopic survey \citep{ubl19} show a weaker trend of $\sigma_{\rm gas}$ with redshift than previously established, with $z\sim 2$ galaxies featuring approximately twice the velocity dispersion observed in local galaxies. Furthermore, $\sigma_{\rm gas}$ in ionized gas, as traced by H$\alpha$, is found to be larger than that of cold molecular gas. This property was recently highlighted by \cite{gir21} who found that galaxies probed with H$\alpha$ data in the DYNAMO and EDGE-CALIFA surveys have $\sim 2-3$ times higher $\sigma_{\rm gas}$ than the cold gas component \citep[see also][]{cor17}. {\it Contemporary high-resolution numerical simulations of turbulent galactic disks confirm this picture of the massive cold phase having much lower intrinsic dispersion than the warm phase} \citep[][]{ejd22,rat22}.

As such, cold gas in high redshift disks, which dominates the ISM in terms of mass, may not necessarily be as turbulent as previously thought
\citep[see also $z>4$ results using ALMA;][]{riz22w,roman23}. This fact aids the development of bars and spiral structure. Furthermore, it is important to note that the presence of highly turbulent, ionized gas does not necessarily affect bar or spiral structure development; such a low-mass component will be dynamically decoupled from the colder stellar and molecular disk components, which jointly dictate the level of gravitational stability in the disk\footnote{The key parameters are here $\Sigma_{\rm gas}/\Sigma_{\star}$ and $\sigma_{\rm gas}/\sigma_{\star}$. If the former is small and the latter is large, gas decouples from the stars in terms of its gravitional stability properties, and the stars evolves closer to a single component.} \citep{jog84,rom13}.
\cite{van22} demonstrated, using simulations of gas-rich isolated disk galaxies, that young stars inherit the velocity dispersion of the dense gas from which they  form. The authors found that even for $f_{\rm gas}>50\%$, $\sigma_\star$ would not exceed $\sim 30$ km s$^{-1}$ \citep[see also][]{ejd22}. \cite{gho22} find that bar formation in such thick and hot stellar disks is possible, with bars in thick disks following the overall growth and temporal evolution of the thin disk's bar. However,  the bar in the thick disk is found to be weaker than the bar in the thin disk. 

In follow-up work, we study bar formation in gas-rich, turbulent disk galaxies. While previous work using such models did not feature bars
\citep[e.g.][]{ren21,van22}, those models were not run for more than a few orbital times, and did not reproduce the large disk mass fraction observed at $z=1-3$ \citep{Price2021}, a property we have demonstrated to be key for rapid bar formation. Gas-rich disks feature different gravitational stability properties compare to gas poor galaxies, with massive clumps developing for $f_{\rm gas}>20-30\%$ \citep[e.g.][]{ren21}. HST observations \citep{elm09} have revealed such clumps to be kiloparsecs in size and as massive as $10^9~M_\odot$, which could be detrimental for the development of spiral structure and bars. However, JWST observations of lensed galaxies challenge this picture \citep{cla22}, with maximum stellar clump masses found to be significantly smaller \citep[$\sim 10^7M_\odot$; see also][]{des19}. Regardless, even in the presence of massive clumps, it remains to be shown whether gas-rich, turbulent galaxies can rapidly develop bars in the limit of high disk mass fractions, as found for pure stellar disks. This is the topic of our next paper.

\section{Summary remarks and next steps} \label{s:conc}

In our new study, we have set out to investigate and extend important work \citep{fuj18a,fujii19} that deserves wider attention in light of new results for high-redshift disks in recent years \citep[e.g.][]{Price2021,guo22}. We refer to the exponential dependence of the bar formation time ($\tau_{\rm bar}$) as a function of $f_{\rm disk}$ as the `Fujii relation.' 
Above some limiting value of $f_{\rm disk}$ ($\approx 0.30\pm 0.05$),
the bar formation time scales exponentially fast, with $1 < \tau_{\rm bar} < 2$ Gyr for most models. 
In his original paper, Fujii defines the bar as existing once the normalised quadrupole moment has grown to $A_2/A_0 = 0.2$.
This relation is a challenge to simulators of galaxy formation $-$ models with inadequate resolution fall off this curve. Modern cosmological simulations are able to produce early disks with bars and spiral arms (cf. $\S 1$; see also
\citealt[][]{grand17}). 
Bars can be produced through internal processes or external perturbations - both of these come down to the responsiveness of the disk, which is largely determined by $f_{\rm disk}$ (Tepper-Garc\'ia et al 2023). But if the bar/spiral arm onset times are representative of the instability timescale rather than an artefact of resolution, these simulations must fall on or close to the Fujii relation.

The Fujii relation is slightly modified (Fig.~\ref{f:taubarexp}) if a more physically motivated definition of bar formation is used ($\S 3$). For the first time, we exploit the exponential growth timescale for $A_2/A_0$ associated with a positive feedback cycle as the bar emerges from the underlying disk. This leads us to a relation that is mass dependent  where the fitting formulae are given in Eq.~\ref{e:fit2}-\ref{e:fit4}. The mass range is appropriate for a Milky Way precursor in the redshift interval $z=1-5$.

Our study suggests that the presence of a bar in a high-redshift disk puts a lower limit on $f_{\rm disk}$ for a given redshift. Fig.~\ref{f:price} (right) is particularly useful because $f_{\rm disk}$ can be estimated independently from the disk kinematics, as has been demonstrated by the SINFONI team \citep{gen20,forster2020,Price2021}. 
This same figure can be used to estimate the disk formation epoch (i.e. $z_{\rm disk}$) if the parameter space is well sampled. But this would require a deep imaging survey of adequate sensitivity extending to higher redshift ($z > 3$), which has now become possible. For example, the JWST Advanced Deep Extragalactic Survey (JADES) will go significantly deeper than existing NIRCAM imaging targetting the same HST ultra-deep and Great Observatories Origins Deep Survey (GOODS) fields \citep[e.g.][]{bunk21}.

The surprising and compelling new evidence (\S 1) for well-developed disks (stars$+$gas) at early times (up to at least $z\approx 6$) is one of the most pressing problems in galaxy formation studies. As far as we know, this development was not foreseen by contemporary theoretical work.
Interestingly, two decades ago, the first hydro/N-body cosmological simulations tended to overload the central regions of galaxies with baryons. This was known as the ``overcooling problem'' and led to the introduction of feedback mechanisms to disperse the baryons over larger radial scales \citep[e.g.][]{ben10a}. Just how the models are to be fixed to ensure dominant central baryons with net rotation at early times is not at all clear \citep{kre21,gur22}.

We have made the case for tracing departures from axisymmetry as evidence for the importance of baryons over dark matter in the inner regions of disk galaxies. This goes to the heart of how galaxies form and evolve at early times. Stellar bars and spiral arms are now seen well beyond $z\sim 2$; this was quite unexpected. Adding to the mystery, these manifestations were thought to be unlikely to occur in the presence of strong gas turbulence.  {\it In essence, the existence of bars and spiral arms in high-redshift disks limits the amount of ongoing feedback and turbulence that is allowed at these early times.}
It is imperative that we understand the nature and origin of these early disk systems. 

In our next study, the gas fraction and turbulent energy input are varied within the HBD framework in order to understand their impact on the disk stellar dynamics and the onset of low-order modes that lead to a bar. This ensures that the simulated galaxy is evolved under controlled conditions at high numerical and intrinsic spatial resolution.

A complementary approach is to run cosmological simulations that track the merger history of an evolving galaxy disk. Within these simulations, disk galaxies in a given mass range can be identified and run again at higher resolution. Some of these disks show signs of high gas fractions, high star formation and turbulence rates. Other disks have evolved stellar populations with essentially no signs of ongoing star formation. But at the present time, such simulations are {\it not} able to account for the discovery of early galactic disks.

\section{Acknowledgments}
JBH wishes to thank Professors Francois Boulanger and Nick Kaiser for hosting him at the \'Ecole Normale Sup\'erieure during the final stages of this work. JBH would also like to thank Edith Falgarone, Fran\c{c}oise Combes, Joe Silk, Paola Di Matteo and Misha Heywood for insightful conversations. JBH and KCF acknowledge an ARC Discovery Project grant (2022-25) that partly supports this work. TTG acknowledges partial financial support from the Australian Research Council (ARC) through an Australian Laureate Fellowship awarded to JBH. OA acknowledges support from the Knut and Alice Wallenberg Foundation, and from the Swedish Research Council (grant 2019-04659). This work was made possible by the National Computing Infrastructure (NCI) Adapter Scheme, with computational resources provided by NCI Australia, an NCRIS-enabled capability supported by the Australian Government. Finally, we are indebted to an insightful referee who encouraged us to think harder about the implications of this work, in addition to improving our overall presentation.

\appendix
\section{Efstathiou, Lake \& Negroponte criterion}

\cite{efs82} derived a simple criterion for bar instability based on a disk's mass $M_{\rm disk}$, scalelength $R_{\rm disk}$ and maximum rotation velocity $V_{\rm max}$, such that for
\begin{equation}
    \varepsilon = V_{\rm max} / (G M_{\rm disk} / R_{\rm disk})^{0.5}
    \label{e:eln}
\end{equation}
then the disk is bar unstable when $\varepsilon \lesssim  1.1$ and stable otherwise. They arrived at the formula from 2D stellar disk simulations held within a rigid halo. Subsequently, \cite{chr95} derived a similar relation for purely gas disks. \cite{sel86} and \cite{ath08} have exposed shortcomings in the use of the ELN relation \citep[see also][]{rom23}: specifically, ELN does not accommodate the contribution of the stellar dispersion or a central bulge, if it exists.
We note, however, that the ELN criterion is still popular among cosmological N-body simulators, regardless of these shortcomings \citep[e.g.][]{izq22}. 

\begin{figure*}[!htb]
    \centering
    \includegraphics[width=0.7\textwidth]{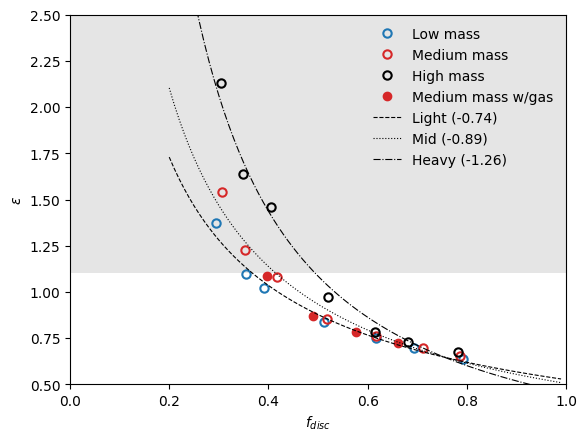}
    \caption{The ELN criterion $\varepsilon$ vs. disk mass fraction $f_{\rm disk}$ for our simulated disks. 
    The shaded area indicates bar models that are stable against bar formation according to the ELN criterion. The curves correspond -- for each halo mass -- to a functional dependence in the form of a power-law, \mbox{$\varepsilon \propto \left(f_{\rm disk}\right)^\alpha$}; the value of the index $\alpha$ is indicated next to each corresponding label. Note the weak dependence of the criterion on the halo mass $M_{\rm halo}$. Note that we do not fit the gas-bearing models, to avoid confusion in the figure: these points are not statistically different from the mid-mass points without gas.
    }
    \label{f:eln}
\end{figure*}

In Fig.~\ref{f:eln}, the ELN parameter is presented for all of our models. A comparison with ~Fig.~\ref{f:taubarexp} shows that there are a number of models that do form a bar in our numerical experiments that would be considered stable based on the ELN criterion (diagonally shaded region). Thus we concur with \citet{ath08} in that the latter is not a reliable estimator of a disk's stability against bar formation.

In view of the definitions of $f_{\rm disk}$ and $\varepsilon$, we expect an inversely proportional relation between these quantities. The simplest and at the same time most general relation is a power law \mbox{$\varepsilon \propto \left(f_{\rm disk}\right)^\alpha$}, with $\alpha < 0$. We have fitted a power-law function to each of the results for a given halo mass model, and find that it provides a reasonable inverse relation between $\varepsilon$ and $f_{\rm disk}$, although $\alpha$ has a secondary dependence on halo mass (cf. Fig~\ref{f:eln}).

\bibliography{main}{}
\bibliographystyle{aasjournal}

\end{document}